\documentclass[12pt,preprint]{aastex}
\usepackage{amsmath,bm}
\usepackage{multirow}
\usepackage{enumerate}
\usepackage{graphicx}
\usepackage{color}
\usepackage[T1]{fontenc}
\usepackage{url}
\usepackage{hyperref}
\newcount\doepsf
\newcommand{\be}{\begin{equation}}
\newcommand{\ee}{\end{equation}}
\newcommand{\bea}{\begin{eqnarray}}
\newcommand{\eea}{\end{eqnarray}}
\newcommand{\bean}{\begin{eqnarray*}}
\newcommand{\eean}{\end{eqnarray*}}

 \definecolor{DarkGreen}{rgb}{0.0,0.45,0.0}     
 \definecolor{DarkMagenta}{rgb}{0.45,0.0,0.45}  

\begin{document}

\title{Magnetic reconnection in strongly magnetized regions of the low solar chromosphere}

\author{Lei Ni$^{1,2}$,
             Vyacheslav S. Lukin$^{3}$ \footnote{Any opinion, findings, and conclusions or recommendations expressed in this material are those of the authors and do not necessarily reflect the views of the National Science Foundation},
            Nicholas A. \ Murphy$^{4}$and
            Jun Lin $^{1,2}$ 
            }

\affil{$^1$Yunnan Observatories, Chinese Academy of Sciences, P. O. Box 110, Kunming, Yunnan 650216, P. R. China}
\affil{$^2$Center for Astronomical Mega-Science, Chinese Academy of Sciences, 20A Datun Road, Chaoyang District, Beijing 100012, P. R. China}
\affil{$^3$National Science Foundation, 2415 Eisenhower Ave, Alexandria, VA 22314, USA} 
\affil{$^4$Harvard-Smithsonian Center for Astrophysics, 60 Garden Street, Cambridge, Massachusetts 02138, USA}

\shorttitle{Magnetic Reconnection in the temperature minimum region}  
\shortauthors{Ni et al.}

\email{leini@ynao.ac.cn}

\slugcomment{MR in TMR; v.\ \today}

\begin{abstract}
\noindent  Magnetic reconnection in strongly magnetized regions around the temperature minimum region of the low solar atmosphere is studied by employing MHD-based simulations of a partially ionized plasma within a reactive 2.5D multi-fluid model. It is shown that in the absence of magnetic nulls in a low $\beta$ plasma the ionized and neutral fluid flows are well-coupled throughout the reconnection region.  However, non-equilibrium ionization-recombination dynamics play a critical role in determining the structure of the reconnection region, lead to much lower temperature increases and a faster magnetic reconnection rate as compared to simulations that assume plasma to be in ionization-recombination equilibrium. The rate of ionization of the neutral component of the plasma is always faster than recombination within the current sheet region even when the initial plasma $\beta$ is as high as $\beta_0=1.46$. When the reconnecting magnetic field is in excess of a kilogauss and the plasma $\beta$ is lower than 0.0145, the initially weakly ionized plasmas can become fully ionized within the reconnection region and the current sheet can be strongly heated to above $2.5\times10^4$~K, even as most of the collisionally dissipated magnetic energy is radiated away. The Hall effect increases the reconnection rate slightly, but in the absence of magnetic nulls it does not result in significant asymmetries or change the characteristics of the reconnection current sheet down to meter scales. 
\end{abstract}   

\keywords{(magnetohydrodynamics) MHD-- 
           methods: numerical-- 
          magnetic reconnection -- 
          Sun: chromosphere}

\section{Introduction}
\label{s:introduction}
  Magnetic reconnection is a universally important physical process in magnetized plasmas which can change magnetic topology and allows for conversion of magnetic energy into plasma particle and photon energy \citep{2009ARA&A..47..291Z}.  Many astrophysical objects and environments are composed of partially ionized magnetized plasmas. The ionization  degree in the warm neutral interstellar medium is of order $10^{-2}$, and it can be as low as $10^{-7}$ in the dense cold interstellar clouds \citep{2011PhPl...18k1211Z}. In the low solar atmosphere, the plasma density varies sharply with height and the ionization degree of hydrogen also varies from $10^{-4}$ in the photosphere to $1$ at the top of the chromosphere \citep{1981ApJS...45..635V, 1993ApJ...406..319F}. Dynamical events, such as chromospheric jets \citep[e.g.,][]{2009ApJ...707L..37L, 2012A&A...543A...6M, 2013A&A...552L...1B, 2014Sci...346A.315T}, Ellerman Bombs \citep[e.g.,][]{2006ApJ...643.1325F, 2014ApJ...792...13H, 2015ApJ...798...19N} and type II white light flares \citep[e.g.,][]{1994SoPh..149..143D, 1999ApJ...512..454D}, are all related to magnetic reconnection in the partially ionized low atmosphere. Improvements in the angular resolution of solar telescopes have allowed many small scale magnetic reconnection events in the low solar atmosphere to be observed \citep[e.g.,][]{2015ApJ...798L..11Y, 2016NatCo...711837X, 2017ApJ...836...52Z}. The high temperature compact bright points which have UV counterparts that are frequently observed with the Interface Region Imaging Spectrograph (IRIS) \citep{2014Sci...346C.315P, 2015ApJ...812...11V, 2016A&A...593A..32G, 2016ApJ...824...96T} are known as IRIS bombs. They share some characteristics in common with Ellerman Bombs (EBs), e.g., similar life time (about 3-5 min) and size (about $0.3^{\prime \prime}$-$0.8^{\prime \prime}$). However, the temperature of the IRIS bombs identified in Si IV slit-jaws is considered to be an order of  magnitude higher than traditional EBs ($\lesssim 10^4$~K).  Emission in the Si IV 139.3 nm line requires temperature of at least $2\times10^4$~K in the dense photosphere or $(6-8)\times10^4$~K from the upper chromosphere  to the corona. Some high temperature IRIS bombs might be caused by small flaring arch filaments in the upper chromosphere or the transition region\cite{2015ApJ...812...11V, 2016A&A...593A..32G}. Some of these events identified in Si IV slit-jaws  are believed to be generated by magnetic reconnection in the temperature minimum region (TMR) or even in the photosphere \citep{2015ApJ...812...11V, 2016A&A...593A..32G, 2016ApJ...824...96T}.  
   
A currently controversial issue is how high a temperature the plasma in a reconnection process around TMR can be heated to. \citep{2017RAA....17...31F} performed detailed non-LTE calculations of the H${\alpha}$ and Ca II 8542Å line profiles, as well as continuum emission, for three EB models with different temperatures around the TMR. Their semi-empirical modeling showed that a higher temperature (higher than $10^4$~K) was not compatible with observed H${\alpha}$ and Ca II 8542Å line profiles and that higher temperatures would be inconsistent with observations. Many numerical simulations \citep[e.g.,][]{2001ChJAA...1..176C, 2007ApJ...657L..53I, 2009A&A...508.1469A, 2011RAA....11..225X} showed that the maximum temperature increases in magnetic reconnection below the upper chromosphere is always only several thousand K. However, the work by \citet{2016ApJ...832..195N} showed that the plasma can be heated from $4200$~K to above $8\times10^4$~K during magnetic reconnection in the TMR with strong magnetic field and low plasma $\beta$ \citep{2016ApJ...832..195N}. These simulations included ambipolar diffusion, temperature dependent magnetic diffusion, heat conduction, the optically thin radiative cooling deduced from observations \citep{1990ApJ...358..328G} and a heating term, but the plasma was assumed to be in a steady ionization equilibrium state.

Recently, \citep{2017ApJ...839...22H} performed single fluid 3D MHD simulations with radiative transport to simulate EBs and flares  at the surface and the lower atmosphere of the Sun. In their simulations, the reconnection event occuring around the middle chromosphere ($z=1-2$~Mm above the photosphere) can heat the plasma to $\sim7.5\times10^4$~K which leads to the appearance of UV bursts. At the same time, the plasma temperature was observed to increase to only around $10^4$~K during EB formation in the photosphere. However, non-equilibrium ionization effects were not considered in their model. The grid size in their simulations was $20$~km which is at least two orders of  magnitude larger than the neutral-ion collision mean free path in the low solar atmosphere. Therefore, the possibility of decoupling between the ionized and neutral fluids around a reconnection site was not captured, and the artificial hyper-diffusivity operator that was included to prevent the collapse of the current sheets leaves open the possibility of smaller scale and hotter structures at spatial scales not covered in that simulation.

The neutral particles in partially ionized plasmas affect  the reconnection process in several different ways. The magnetic diffusion coefficient in partially ionized plasmas includes contributions from both electron-ion and electron-neutral collisions. Ambipolar diffusion allows for additional decoupling between magnetic fields and the bulk plasma fluid due to a finite neutral-ion collision frequency. Previous theoretical and numerical work \citep{1994ApJ...427L..91B, 1995ApJ...448..734B, 1999ApJ...511..193V, 2015ApJ...799...79N} showed that the current sheet thins rapidly due to ambipolar diffusion when no guide field is present, but even a small guide field will suppress the effect of ambipolar diffusion. The two-fluid model (neutral-ion) has been used to study the magnetic reconnection in partially ionized plasma \citep[e.g.,][]{2006ApJ...642.1236S, 2008ApJ...687L.127S, 2009ApJ...691L..45S}, with ambipolar diffusion naturally included. \citep{2008ApJ...687L.127S} found that the reconnection rate in the upper chromosphere was 20 times larger than in the lower chromosphere, which has a lower ionization degree. However, the ionization and recombination rates were assumed as fixed values rather than depended on temperature and density, so that the important role of recombination in chromospheric reconnection was not captured in that paper. 

\cite{2012ApJ...760..109L, 2013PhPl...20f1202L} used the reactive multi-fluid plasma-neutral module within the HiFi modeling framework \citep{Lukin2016} to study null-point magnetic reconnection in the solar chromosphere. A similar plasma-neutral module within a different code \citep{2017ApJ...842..117A} has been used to investigate the role of radiative cooling on chromospheric reconnection. They found that neutral and ion fluids can become decoupled upstream of the reconnection current sheet but are well-coupled in the outflows. In their work, strong ion recombination in the reconnection region, combined with Alfv\'enic outflows, lead to a fast reconnection rate independent of Lundquist number. \cite{2015ApJ...805..134M} used the same model implementation within the HiFi framework to study asymmetric magnetic reconnection in weakly ionized chromospheric plasmas. The upstream plasma $\beta$ in these previous simulations \citep{2012ApJ...760..109L, 2013PhPl...20f1202L,2015ApJ...805..134M,2017ApJ...842..117A} is greater than 1. In the paper by \cite{ 2013PhPl...20f1202L}, the ionization degree $f_i=n_i/(n_i+n_n)$ within the current sheet is shown to increase by an order of magnitude during the reconnection process, but the highest ionization degree ($f_i=1.2\%$) is still low and the plasma is still weakly ionized throughout the whole magnetic reconnection process. The temperature increase is not significant even after secondary islands appear in these simulations.

On scales comparable to or less than the ion inertial length $d_i$,  the Hall effect is expected to be important in magnetic reconnection dynamics \citep[e.g.,][]{1994GeoRL..21...73M, 2001PhRvL..87z5003W, 2003PhPl...10.3131L, 2005PhRvL..95e5003R, 2006PhPl...13e2119Y}. In partially ionized plasmas with strong ion-neutral coupling, the effective ion inertial length has been predicted to be enhanced as $d_i^{\prime}=d_i\sqrt{\rho/\rho_i}$, and the Hall effect has been predicted to accelerate the reconnection rate for current sheets narrower than $d_i^{\prime}$ \citep{2011ApJ...739...72M}. However, no such reconnection rate acceleration was observed in the null-point reconnection simulations by \cite{2015ApJ...805..134M}, even though signatures of the Hall effect-generated magnetic fields were clearly evident at $d_i^{\prime}$ spatial scales. This difference might be related to the decoupling of the plasma and neutral inflows, whereas the Hall effect enhancement requires strong coupling.  

In this work, we present results and analysis of the first reactive multi-fluid simulations of magnetic reconnection in low $\beta$ plasmas with guide field. The plasma parameters in our simulations are representative of the temperature minimum region in the solar atmosphere. How the plasma $\beta$ and the Hall term affect the reconnection process in initially weakly ionized plasmas will be presented. Section \ref{s:model} describes our numerical model and simulation setup.  We present our numerical results in Section \ref{s:results}.  A summary and discussion are given in Section \ref{s:discussion}.


\section{Numerical model and initial conditions}
\label{s:model}
\subsection{Ionization, Recombination and Charge exchange}
The HiFi module for partially ionized plasmas includes the electron-impact ionization, radiative recombination and resonant charge exchange (CX) interactions \citep{2011PhDthesis, 2012PhPl...19g2508M, 2012ApJ...760..109L, 2013PhPl...20f1202L}. We only consider the hydrogen gas in this work. The subscripts $``n"$, $``i"$ and $``e"$ refer to neutrals, ions and electrons in this work, respectively. The ionization and recombination rates are given by
\begin{equation}
   \Gamma_{n}^{\mathrm{ion}} \equiv -n_n \nu^{\mathrm{ion}},
\end{equation}
\begin{equation}
  \Gamma_i^{\mathrm{rec}} \equiv -n_i \nu^{\mathrm{rec}},
\end{equation}
with $\Gamma_i^{\mathrm{ion}}=-\Gamma_n^{\mathrm{ion}}$ and $\Gamma_i^{\mathrm{rec}}=-\Gamma_n^{\mathrm{rec}}$. The ionization frequency 
 \begin{equation}
   \nu^{\mathrm{ion}} = \frac{n_e A}{X+\phi_{\mathrm{ion}}/T_e^{\ast}}\left(\frac{\phi_{\mathrm{ion}}}{T_e^{\ast}}\right)^{K} \mathrm{exp}(-\frac{\phi_{\mathrm{ion}}}{T_e^{\ast}}) ~~\mbox{m$^{3}$\,s$^{-1}$}
 \end{equation}
 is given by the practical fit from \cite{1997ADNDT..65....1V}, using the values $A=2.91\times10^{-14}$, $K=0.39$, $X=0.232$, and the hydrogen ionization potential $\phi_{\mathrm{ion}}=13.6$~eV. The recombination frequency obtained from \cite{2003poai.book.....S} is
 \begin{equation}
   \nu^{\mathrm{rec}} = 2.6\times10^{-19} \frac{n_e}{\sqrt{T_e^{\ast}}} ~~\mbox{m$^{3}$\,s$^{-1}$}.
 \end{equation}
 
 The CX reaction rate, $\Gamma^{\mathrm{cx}}$is defined as 
 \begin{equation}
    \Gamma^{\mathrm{cx}} \equiv \sigma_{\mathrm{cx}}(V_{\mathrm{cx}})n_in_nV_{\mathrm{cx}},
 \end{equation}
 where
  \begin{equation}
      V_{\mathrm{cx}} \equiv \sqrt{\frac{4}{\pi}V_{Ti}^2+\frac{4}{\pi}V_{Tn}^2+V_{in}^2}
  \end{equation}
  is the representative speed of the interaction and $V_{in}^2 \equiv |\mathbf{V}_i-\mathbf{V}_n|^2$ with $\mathbf{V}_{\alpha}$ denoting the velocity of species $\alpha$. The thermal speed of species $\alpha$ is $V_{T\alpha}=\sqrt{2k_BT_{\alpha}/m_{\alpha}}$, where $T_{\alpha}$ is the temperature, $m_{\alpha}$ is the mass of corresponding particle, and $k_B$ is Boltzmann's constant. The expression for the CX cross-section $\sigma_{\mathrm{cx}}(V_{\mathrm{cx}})$ can be found in \cite{2011PhDthesis} and \cite{2012PhPl...19g2508M}.   
 
\subsection{Model equations}
The equations solved by the plasma-neutral module of HiFi are the same as those in \cite{2015ApJ...805..134M}. The ion and neutral continuity equations are  
\begin{equation}
  \frac{\partial n_i}{\partial t} + \nabla \cdot (n_i \mathbf{V}_i) = \Gamma_i^{\mathrm{rec}} +\Gamma_i^{\mathrm{ion}},  
\end{equation}

\begin{equation}
   \frac{\partial n_n}{\partial t}+\nabla \cdot (n_n \mathbf{V}_n)=\Gamma_n^{\mathrm{rec}}+\Gamma_n^{\mathrm{ion}}.
\end{equation}
We assume $n_i=n_e$ in these equations. 

The ion and neutral momentum equations are given by
\begin{eqnarray}
   \frac{\partial}{\partial t} (m_in_i\mathbf{V}_i)+\nabla \cdot (m_in_i\mathbf{V}_i\mathbf{V}_i+\Bbb P_i+\Bbb P_e)  \nonumber \\
                                       = \mathbf{J} \times \mathbf{B} + \mathbf{R}_i^{\mathrm{in}}+                                                      
                                            \Gamma_i^{\mathrm{ion}}m_i\mathbf{V}_n-\Gamma_n^{\mathrm{rec}}m_i\mathbf{V}_i   \nonumber \\
                                           + \Gamma^{\mathrm{cx}}m_i(\mathbf{V}_n-\mathbf{V}_i)+                                    
                                               \mathbf{R}_{\mathrm{in}}^{\mathrm{cx}}-\mathbf{R}_{\mathrm{ni}}^{\mathrm{cx}}, \\ 
   \frac{\partial}{\partial t} (m_in_n\mathbf{V}_n)+\nabla \cdot (m_in_n\mathbf{V}_n\mathbf{V}_n+\Bbb P_n)  \nonumber \\
                                       = -\mathbf{R}_i^{in}+                                                      
                                            \Gamma_n^{\mathrm{rec}}m_i\mathbf{V}_i-\Gamma_i^{\mathrm{ion}}m_i\mathbf{V}_n   \nonumber \\
                                           + \Gamma^{\mathrm{cx}}m_i(\mathbf{V}_i-\mathbf{V}_n)-                                    
                                               \mathbf{R}_{\mathrm{in}}^{\mathrm{cx}}+\mathbf{R}_{\mathrm{ni}}^{\mathrm{cx}}.  
\end{eqnarray}
Here the current density is 
\begin{equation}
   \mathbf{J}=en_i(\mathbf{V}_i-\mathbf{V}_e)=\frac{\nabla \times \mathbf{B}}{\mu_0}
\end{equation}
Note that the Lorenz force $\mathbf{J}\times\mathbf{B}$ only acts on the plasma but not the neutrals. 
The momentum transfer $\mathbf{R}_\alpha^{\alpha \beta}$ is the transfer of momentum to species $\alpha$ due to identity-preserving collisions with species $\beta$:
\begin{equation}
  \mathbf{R}_\alpha^{\alpha \beta}=m_{\alpha \beta} n_{\alpha} \nu_{\alpha \beta}(\mathbf{V}_{\beta}-\mathbf{V}_{\alpha}),
\end{equation}
where $m_{\alpha \beta}=m_{\alpha}m_{\beta}/(m_{\alpha}+m_{\beta})$. The collision frequency $\nu_{\alpha \beta}$ is 
\begin{equation}
  \nu_{\alpha \beta} = n_{\beta} \Sigma_{\alpha \beta} \sqrt{\frac{8k_BT_{\alpha \beta}}{\pi m_{\alpha \beta}}},
\end{equation}
with $T_{\alpha \beta}=(T_{\alpha}+T_{\beta})/2$. We choose the cross-section $\Sigma_{in}=\Sigma_{ni}=5\times10^{-19}$ m$^{-2}$ as in \cite{2013PhPl...20f1202L} and \cite{2015ApJ...805..134M}. The momentum transfer from species $\beta$ to species $\alpha$ due to CX is $\mathbf{R}_{\alpha \beta}^{\mathrm{cx}}$, and the appropriate approximations for these terms between ions and neutrals as presented in \cite{2012ApJ...760..109L} are 
\begin{equation}
  \mathbf{R}_{\mathrm{in}}^{\mathrm{cx}} \approx -m_i\sigma_{\mathrm{cx}}(V_{\mathrm{cx}})n_in_n\mathbf{V}_{\mathrm{in}}V_{Tn}^2\left[  
                  4\left( \frac{4}{\pi}V_{Ti}^2+V_{in}^2\right)+\frac{9\pi}{4}V_{Tn}^2\right]^{-1/2}
\end{equation}
and
\begin{equation}
  \mathbf{R}_{\mathrm{ni}}^{\mathrm{cx}} \approx m_i\sigma_{\mathrm{cx}}(V_{\mathrm{cx}})n_in_n\mathbf{V}_{\mathrm{in}}V_{Ti}^2\left[  
                  4\left( \frac{4}{\pi}V_{Tn}^2+V_{\mathrm{in}}^2\right)+\frac{9\pi}{4}V_{Ti}^2\right]^{-1/2}.
\end{equation}
The pressure tensor for species $\alpha$ is $\Bbb P_{\alpha}=P_{\alpha} \mathrm{\Bbb I}+\pi_{\alpha}$ where $P_{\alpha}$ is the scalar pressure and $\pi_{\alpha}$ is the viscous stress tensor, given by $\pi_{\alpha}=-\xi_{\alpha}\left[ \nabla \mathbf{V}_{\alpha}+(\nabla \mathbf{V}_{\alpha})^{\top}\right]$ with $\xi_{\alpha}$ as the isotropic dynamic viscosity coefficient. 

Combining the electron and ion energy equation together and neglecting terms of order $(m_i/m_p)^{1/2}$ and higher, one can get:
\begin{eqnarray}
  \frac{\partial}{\partial t}(\varepsilon_i+\frac{P_e}{\gamma-1}) 
                             +   \nabla \cdot (\varepsilon_i\mathbf{V}_i+\frac{P_e\mathbf{V}_e}{\gamma-1}+
   \mathbf{V}_i \cdot \Bbb P_i+\mathbf{V}_e \cdot \Bbb P_e+ \mathbf{h}_i+\mathbf{h}_e)  \nonumber \\
   = \mathbf{j} \cdot \mathbf{E} + \mathbf{V}_i \cdot \mathbf{R}_i^{\mathrm{in}}+Q_i^{\mathrm{in}} -
      \Gamma_n^{\mathrm{rec}}\frac{1}{2}m_i\mathbf{V}_i^2-Q_n^{\mathrm{rec}}+ \nonumber \\
      \Gamma_i^{\mathrm{ion}}\left(\frac{1}{2}m_i\mathbf{V}_n^2-\phi_{\mathrm{eff}}\right)+
      Q_i^{\mathrm{ion}}+\Gamma^{\mathrm{cx}}\frac{1}{2}m_i\left( \mathbf{V}_n^2-\mathbf{V}_i^2\right)+ \nonumber \\
      \mathbf{V}_n \cdot \mathbf{R}_{\mathrm{in}}^{\mathrm{cx}}-\mathbf{V}_i \cdot \mathbf{R}_{\mathrm{ni}}^{\mathrm{cx}}
     + Q_{\mathrm{in}}^{\mathrm{cx}}-Q_{\mathrm{ni}}^{\mathrm{cx}}. 
\end{eqnarray} 

\begin{eqnarray}
  \frac{\partial \varepsilon_n}{\partial t} 
                             +   \nabla \cdot (\varepsilon_n\mathbf{V}_n+
   \mathbf{V}_n \cdot \Bbb P_n+ \mathbf{h}_n)  \nonumber \\
   = -\mathbf{V}_n \cdot \mathbf{R}_i^{\mathrm{in}}+Q_n^{\mathrm{ni}} -
      \Gamma_i^{\mathrm{ion}}\frac{1}{2}m_i\mathbf{V}_n^2-Q_i^{\mathrm{ion}}+ \nonumber \\
       \Gamma_n^{\mathrm{rec}}\frac{1}{2}m_i\mathbf{V}_i^2+ 
      Q_n^{\mathrm{rec}}+\Gamma^{\mathrm{cx}}\frac{1}{2}m_i\left( \mathbf{V}_i^2-\mathbf{V}_n^2\right)+ \nonumber \\ 
      \mathbf{V}_i \cdot \mathbf{R}_{\mathrm{ni}}^{\mathrm{cx}}-\mathbf{V}_n \cdot \mathbf{R}_{\mathrm{in}}^{\mathrm{cx}}+ 
      Q_{\mathrm{ni}}^{\mathrm{cx}}-Q_{\mathrm{in}}^{\mathrm{cx}}. 
\end{eqnarray} 
The internal energy density is $\varepsilon_{\alpha} \equiv m_{\alpha}n_{\alpha}\mathbf{V}_{\alpha}^2/2+P_{\alpha}/(\gamma-1)$. The term $\Gamma_i^{\mathrm{ion}} \phi_{\mathrm{eff}}$ represents assumed optically thin radiative losses with $\phi_{\mathrm{eff}}=33$~eV \citep{2012ApJ...760..109L}. The ratio of specific heats is denoted by $\gamma$. $Q_{\alpha}^{\alpha \beta}$ represents the heating of species $\alpha$ due to interaction with species $\beta$, which is a combination of frictional heating and a thermal transfer between the two populations: $Q_{\alpha}^{\alpha \beta}=\mathbf{R}_{\alpha}^{\alpha \beta} \cdot (\mathbf{V}_{\beta}-\mathbf{V}_{\alpha})+3m_{\alpha \beta}n_{\alpha}\nu_{\alpha \beta}(T_{\beta}-T_{\alpha})$. The electron and ion heat conduction terms $\mathbf{h}_e$ and $\mathbf{h}_i$ are given by 
\begin{equation}
   \mathbf{h}_{\alpha} = \left[ \kappa_{\parallel, \alpha} \mathbf{\hat{b}} \mathbf{\hat{b}}+\kappa_{\perp, \alpha}(\Bbb I-\mathbf{\hat{b}} \mathbf{\hat{b}}) \right] \cdot \nabla k_{B} T_{\alpha}, 
   \end{equation}
where $\kappa_{\parallel, \alpha}$ and $\kappa_{\perp, \alpha}$ are the conductivity coefficients \citep{1965RvPP....1..205B} which are parallel and perpendicular to the magnetic field direction $\mathbf{\hat{b}}$, respectively. The isotropic neutral heat conduction is $\mathbf{h}_n=-\kappa_n\nabla k_BT_n$. The changes in thermal energy of ionized and neutral plasma components due to ionization or recombination are $Q_i^{\mathrm{ion}}=\Gamma_i^{\mathrm{ion}}(3/2)k_BT_n$ and $Q_n^{\mathrm{rec}}=\Gamma_n^{\mathrm{rec}}(3/2)k_BT_i$. $Q_{\alpha \beta}^{\mathrm{cx}}$ denotes heat flow from species $\beta$ to species $\alpha$ due to CX \citep{2011PhDthesis, 2012PhPl...19g2508M}.

The generalized Ohm's law for this work is given by
\begin{equation}
   \mathbf{E}+\mathbf{V}_i \times \mathbf{B} = \eta \mathbf{J}+\frac{\mathbf{J}\times\mathbf{B}}{en_i}-
   \frac{\nabla \cdot \Bbb P_e}{en_i}-\frac{m_e\nu_{en}}{e}(\mathbf{V}_i-\mathbf{V}_n).
\end{equation}
Here the magnetic diffusion coefficient includes both the ion-electron and electron-neutral collisions and is written as 
\begin{equation}
  \eta = \frac{m_en_e(\nu_{ei}+\nu_{en})}{(en_e)^2}.
\end{equation}

\subsection{Normalizations and Initial conditions}
We normalize the equations by using the characteristic plasma density and magnetic field around the TMR of the low solar chromosphere. The characteristic plasma number density is $n_{\star}=10^{21}$~m$^{-3}$, and the characteristic magnetic field is $B_{\star}=0.05$~T=$500$~G. We focus on small-scale magnetic reconnection processes and choose a characteristic length of $L_{\star}=100$~m. From these quantities, we derive additional normalizing values to be $V_{\star}\equiv B_{\star}/\sqrt{\mu_0m_pn_{\star}}=34.613$~km\,s$^{-1}$, $t_{\star}\equiv L_{\star}/V_{\star}=t_A=0.0029$~s and $T_{\star} \equiv B_{\star}^2/(k_B\mu_0n_{\star})=1.441\times10^5$~K. The initial ionized and neutral fluid densities are set to be uniform with the neutral particle number density of $n_{n0}=0.5n_{\star}=0.5\times10^{21}$~m$^{-3}$, and the initial ionization degree is $f_{i0}=n_{i0}/(n_{i0}+n_{n0})=0.01\%$. Thus, the neutral-ion collisional mean free path of the background plasma is $\lambda_{ni0}=23.74$~m, and the ion inertial length is $d_{i0}=0.99$~m. Since the hydrogen gas only has the ground state and the ionized state in this model, the initial temperatures of the ionized and neutral fluids are set to be uniform at $T_{i0}=T_{n0}=8400$~K to keep the ionization degree the same as that around the TMR in solar atmosphere. 

The dimensionless magnetic diffusion form is 
\begin{equation}
   \eta=\eta_{ei}+\eta_{en}=\eta_{ei\star}T_e^{-1.5}+\eta_{en\star}(T_e+T_n)^{1/2}\frac{n_n}{n_e},
\end{equation}  
where $\eta_{ei\star}=7.457\times10^{-6}$ and $\eta_{en\star}=1.369\times10^{-6}$ are normalization constants derived from the characteristic values $n_{\star}$, $B_{\star}$, and $L_\star$. $T_e$, $T_n$, $n_e$ and $n_n$ are the dimensionless temperatures and number densities for the electron and neutral fluids respectively. Since the electron and the ion are assumed to be coupled together and only the hydrogen gas is considered in our model, we assume $T_i=T_e$, $n_i=n_e$, and the pressure of the ionized component is twice the ion (or electron) pressure, $P_p = P_e + P_i = 2P_i$. The initial magnetic diffusion contribution from electron-neutral collisions is about one magnitude higher than that due to electron-ion collisions. However, as is shown below, magnetic diffusion during reconnection becomes dominated by the electron-ion collisions as the ionization degree increases and $n_n/n_e$ decreases with time inside the current sheet. We have also calculated the Lundquist number $S=LV_A/\eta$ by using the magnetic diffusion, the Alfv\'en speed and the length of the current sheet in our simulations. Since the length scale is very small, the Lundquist number during the magnetic reconnection process is only around $2000$ in this work. In agreement with previous simulations of reconnecting current sheets \citep[e.g.,][]{2017ApJ...849...75H, 2016PhPl...23j0702C, 2013PhPl...20f1206N, 2012PhPl...19g2902N, 2012ApJ...760..109L, 2010PhPl...17f2104H}, no plasmoid instability appears in our simulations.    

The simulations are initialized with a force-free Harris sheet magnetic equilibrium, where the reconnecting magnetic field is along the $y$ coordinate direction and the guide B-field is in the $z$-direction. Specifically, the initial dimensionless magnetic flux in $z$ direction is given by 
\begin{equation}
  A_{z0}(y)=-b_p\lambda_{\psi} \mathrm{ln} \left[\mathrm{cosh}\left(\frac{y}{\lambda_{\psi}}\right)\right],
\end{equation}
where $b_p$ is the strength of the the magnetic field and $\lambda_{\psi}$ is the initial thickness of the current sheet. The initial magnetic field in z-direction is  
\begin{equation}
  B_{z0}(y)=b_p \bigg/  \left[ \mathrm{cosh}\left(\frac{y}{\lambda_{\psi}}\right) \right].
\end{equation}

We have simulated six cases: Case~A, Case~A0, Case~B, Case~C, Case~D and Case~E. In order to see how the non-equilibrium ionization-recombination effect impacts  the reconnection process, we have eliminated this effect in Case~A0. In Case~A0, the recombination rate does not depend on the plasma density and temperature as in equation (2) and (4). Instead, we compel the recombination rate $\Gamma_i^{\mathrm{rec}}$ to equal the ionization rate $\Gamma_n^{\mathrm{ion}}$ at all times, and the right hand sides of equation~(7) and equation (8) are always zero in this case. The non-equilibrium ionization-recombination is included in all the other cases. Except for the non-equilibrium ionization-recombination factor, all the other conditions in Case~A0 are the same as in Case~A. The electron-neutral collisions and the Hall term are only included in Case~B, but the initial conditions in Case~A and Case~B are identical. The only difference among Case~A, Case~C, Case~D and Case~E is the strength of the initial magnetic field: $b_p=1$ in Case~A, $b_p=0.2$ in Case $C$, $b_p=2$ in Case~D and $b_p=3$ in Case~E. Therefore, one can calculate the initial plasma $\beta$ in each case: $\beta_0=0.058$ in Case~A, A0 and B, $\beta_0=1.46$ in Case~C, $\beta_0=0.0145$ in Case~D and $\beta_0=0.0064$ in Case~E. The differences in initial conditions and evolution equations among these six cases are summarized in Table~1. The reconnection processes are symmetric in both x and y direction in Cases~A, A0, C, D and E. We therefore only simulate one quarter of the domain ($0<x<2$, $0<y<1$) in Cases~A, A0, C, D and E. The Hall effect might result the asymmetries \citep{2003PhPl...10.3131L} and we simulate the full domain ($-2<x<2$, $-1<y<1$) in Case~B. We use the same outer boundary conditions at $|y| = 1$ and the same form of the initial electric field perturbations as \cite{2015ApJ...805..134M} to initiate magnetic reconnection in all of the cases in this work. The perturbation electric field is applied for $0\leq t \leq 1$. The perturbation magnitude is proportional to the value of $b_p$ in each of the cases with the amplitude of $\delta E=10^{-3}b_p$. Periodicity of the physical system is imposed in the x-direction at $|x| = 2$.

All the six cases have been tested by using a lower resolution and a higher resolution. The highest number of grid elements in Cases~A, A0, C, D and E is $m_x=64$ elements in the $y$-direction and $m_y=64$ elements in the $x$-direction. The highest number of grid elements in Case~B is $m_x=128$ and $m_y=128$. We use sixth order basis functions for all simulations, resulting in effective total grid size $(M_x, M_y)=6(m_x,m_y)$. Grid packing is used to concentrate mesh in the reconnection region. Therefore, the mesh packing along the y-direction is concentrated to a thin region near y=0. Note that the quantities shown in the figures in this work are in dimensionless units except for temperature and velocity which can be used to compare with the temperature and velocity derived from a model of the line forming process and observations of spectral line profiles.

\section{Numerical Results}\label{s:results}
\subsection{Magnetic reconnection with different plasma $\beta$ in initially weakly ionized plasmas around TMR}\label{ss:MR:I}
 We have simulated magnetic reconnection with different strengths of initial magnetic field, which results in different plasma $\beta$.  How the plasma $\beta$ affects the reconnection process in initially weakly ionized plasmas around TMR will be presented in this subsection.
 
 First, we compare the reconnection process in Case~A with initial $\beta_{0}=0.058$ and Case~C with initial $\beta_{0}=1.46$. Fig.~1 shows the current density $J_z$ and ionization degree $f_i$ in one quarter of the domain at three different times in Case~A and Case~C respectively. The current sheet lengths in Case~A at $t=4.024$ and in Case~C at $t=20.052$ are the same, as are those shown in Fig.~1(b) and Fig.~1(e), and those shown in Fig.~1(c) and Fig.~1(f). As presented in Fig.~1, about $40\%$ of the neutral particles are eventually ionized inside the current sheet region in Case~A. In contrast, while the maximum ionization degree in Case~C is increased during the reconnection process by a factor of 300 from $0.01\%$ to $3\%$, the plasmas are still weakly ionized even during the later stage of the reconnection process. Fig.~3 shows that the neutral fluid can be fully ionized inside the current sheet when the magnetic field is strong enough as in Case~D and Case~E. 

Fig.~2 shows the profiles of ion and neutral temperatures across the current sheet in Cases A and C at the same three pairs of times as in Fig.~1. It is apparent that (1) in both cases, the ion and neutral temperatures are nearly equal throughout the evolution due to rapid thermal exchange between the plasma components; (2) the initial temperature increase observed in Case A due to the Joule heating within the broad Harris current sheet prior to the onset of the reconnection process is later moderated by radiative cooling; and (3) peak temperature within the narrow reconnection current sheet is maintained at about $1.6\times 10^4$~K in both cases. As shown in Fig.~3, the plasma inside the current sheet can only be heated to above $2\times10^4$~K when the plasma becomes fully ionized, the highest temperature can reach around $4.6\times10^4$~K in Case~E. 

 We have calculated the time evolution of the total radiated energy $Q_{rad}=\int_{0}^{1}\int_{0}^{2} L_{rad} dxdy$, the Joule heating $Q_{Joule}=\int_{0}^{1}\int_{0}^{2} \eta J^2 dxdy$, the frictional heating between ions and neutral particles $Q_{in}$ and the viscous heating of ions and neutral particles $Q_{vis}$ in the whole simulation domain. Together, these four terms represent all sources and sinks of plasma thermal energy within the domain. In all simulations described here, both $Q_{in}$ and $Q_{vis}$ are found to be several magnitudes smaller than Joule heating $Q_{Joule}$, (not shown). The time evolution of these quantities for Case A is plotted in Fig. 4(a), showing that the radiated thermal energy strongly increases once the reconnection current sheet is formed. This is consistent with our interpretation of the temperature evolution shown in Fig. 2(a). Further, it is shown that most of the total generated thermal energy $Q_{Joule}+Q_{in}+Q_{vis}$ is radiated during the rapid ionization-recombination cycle in Case~A. In Case~C with high plasma $\beta$ and weak magnetic field, the radiated thermal energy increases much slower with time during the reconnection process as shown in Fig.~4(b). Though, a large amount of the thermal energy is similarly radiated during the later stage of the reconnection process.      

From our simulations results, we find that the ionized and neutral fluids are well coupled in the reconnection outflow regions. This is consistent with the previous results by \cite{2012ApJ...760..109L, 2013PhPl...20f1202L}, and by \cite{2015ApJ...805..134M}.  These prior studies in the high $\beta$ regime also showed that the neutral and ionized fluid components decouple upstream of the reconnection site on scales smaller than $\lambda_{ni}$.  As shown in Fig. 5, similar behavior is observed in our Case C simulation, but the neutral and ion inflows are observed to be well-coupled in the steady-state reconnection phase in Case A, as well as in the other low $\beta$ Cases B, D, and E (not shown). We note that for the low $\beta$ plasma in Case~A, there is a positive peak value of $V_{iy}-V_{ny}$ in the upstream of the reconnection site, the peak value of $V_{iy}-V_{ny}$ is at around $y=0.125$ at $t=20.126$, which means the speed of the inflowing neutral fluid is even greater than that of the ion fluid at this position.

One possible explanation for the observed differences between the high $\beta$ and low $\beta$ cases lies in the degree of plasma ionization self-consistently formed within the reconnection current sheets.  We note that for two plasma elements with the same total atom particle density, a plasma element that is $50\%$ ionized would have the neutral-ion mean free path be~50 times shorter than that which is $1\%$ ionized. Thus, the decoupling between ion and neutral fluid inflows observed in the high $\beta$ simulations that remain weakly ionized can be suppressed in the low $\beta$ simulations where the plasma becomes strongly ionized.

Panels (c) and (f) of Fig. 5 show the terms contributing to the pressure balance across the reconnection current sheet during the quasi-steady-state phase in Cases A and C, respectively.  We note that in both cases neither neutral pressure nor guide-field $B_z$ act to balance the magnetic field pressure from the reconnecting B-field component within the current sheet.  As a result, as in the previous work by \cite{2013PhPl...20f1202L}, the decrease in the magnetic field pressure of the reconnecting field has to be balanced by a corresponding increase in the ionized fluid density and pressure. This observation points to a conclusion that the ionization fraction within a reconnecting current sheet in a partially ionized plasma is primarily controlled by force-balance requirements (compression) rather than by plasma temperature increase due to Ohmic heating.

The maximum values of the inflow plasma velocity are almost the same in Cases~A, C, D and E with different plasma $\beta$, but the outflow velocity is higher in the lower plasma $\beta$ case. The maximum $V_{ix}$ in Case~E is about $16$~km\,s$^{-1}$ and it is about $7$~km\,s$^{-1}$ in Case~C. 

We have used the same method as \cite{2013PhPl...20f1202L} to calculate the reconnection rate,  $M_{sim}=\eta^{\ast} j_{max}/(V_A^{\ast} B_{up})$, where $j_{max}$ is the maximum value of the out of plane current density $J_{z}$, located at $(x,y)=(0,0)$ in all the simulations in this work. $B_{up}$ is $B_x$ evaluated at $(0, \delta_{sim})$, where $\delta_{sim}$ is defined as the half-width at half-max in $J_{z}$. $V_{A}^{\ast}$ is the relevant  Alfv\'en velocity defined using $B_{up}$ and the total number density $n^{\ast}$ at the location of $j_{max}$. $\eta^{\ast}$ is the magnetic diffusion coefficient defined by in Equation (21) at the location of $j_{max}$, where the electron-neutral collisions are only included in Case~B. Fig.~6(a) shows the time evolution of the reconnection rates in Cases~A, B, C and D, and Fig. 6(b) shows the time evolution of the normalized magnetic diffusion coefficient $\eta^{\ast}$ at the X-point. We note that each of the reconnection rate curves includes both the formation phase of the reconnection region and the later quasi-steady-state phase; and the provided reconnection rate measure is most meaningful in the latter well-developed phase of each of the simulations. The measured reconnection rate in Case~C with high $\beta$ is several times higher than that in the other lower $\beta$ cases. Maximum of the reconnection rate reaches above 0.1 in Case~C, while the maximum rate in Case~A, Case~B and Case~D is only around 0.025. We do not presently have a quantitative theory to explain the measured reconnection rates, this result is consistent with a conjecture that decoupling of ionized and neutral fluid inflows, as observed in Case C but not in Cases A, B, or D, can accelerate the reconnection rate within a reconnection current sheet. Fig. 6(b) further shows that the magnetic diffusion coefficient $\eta^{\ast}$ at the reconnection X-point has approximately the same value of $2\times10^{-4}$ in Cases A, B, and C during the quasi-steady-state reconnection phase. Thus, it has no contribution to the difference in the reconnection rates among these three simulations; and further provides the evidence that  omitting the contribution of electron-neutral collisions to the electrical resistivity in Case A, as compared to the Case B, did not significantly modify the outcome. $\eta^{\ast}$ in Case~D drops to a lower value because of the heating at the reconnection X-point during the later stage of the reconnection process. 

Figs.~7(a) and 7(b) show the four terms in Eq. (7) contributing to $\partial n_i/\partial t$ in Cases~A and C. The values of the four contributions are the average values inside the current sheet domain at each time. Fig.~7(c) shows the variations of the half length and width of the currents versus times in Cases~A and C. The half length of the current sheet abruptly drops to $0.5-0.6$ during a very short period in both Cases~A and C, and the half width of the sheet gradually drops to around $0.015$. In each case, the time when the measured length of the current sheet abruptly drops corresponds to the formation of a fully non-linear reconnection current sheet and onset of the self-consistent reconnection process. In Case~A, the ionization is much larger than the other three terms prior to the formation of a reconnection current sheet, but the contribution of all four terms substantially increases with the onset of reconnection. The value of $-\Gamma_i^{rec}+\partial (n_iV_{ix})/\partial x$ gradually gets close to the value of $-\partial (n_i V_{iy})/{\partial y}+\Gamma_i^{ion}$ during the later stage of the reconnection process, and the current sheet ion density reaches an approximate steady state in Case~A. In Case~C, the ionization rate is always the maximum among the four components as shown in Fig.~7(b). However, we should also note that the total value of contributing terms to $\partial n_i/\partial t$ in Case A is about two orders of magnitude larger than in Case C. The imbalance between $-\Gamma_i^{rec}+\partial (n_iV_{ix})/\partial x$ and $-\partial (n_i V_{iy})/{\partial y}+\Gamma_i^{ion}$ in Case~C is small relative to the neutral fluid density. In both \cite{2012ApJ...760..109L} and \cite{2013PhPl...20f1202L} studies of null-point reconnection, the ionization rate $\Gamma_i^{ion}$ did not play an important dynamical role and it was always the minimum among the four components, which is significantly different from the simulation results presented here. The ionization rate $\Gamma_i^{ion}$ always plays an important role in the whole reconnection process in all the cases studied in this work. From Figs.~7 (a) and 7(b), we can see that the ionization rate $\Gamma_i^{ion}$ is  always higher than the recombination rate $-\Gamma_i^{rec}$ in both Cases~A and C. The plasma $\beta$ inside the current sheet in Case~C is still smaller than that in previous work by \cite{2013PhPl...20f1202L} without guide field. Therefore, the plasma $\beta$ and the magnetic field structures appear to determine whether ionization will be dynamically important inside the current sheet region.

\subsection{The non-equilibrium ionization-recombination effect in initially weakly ionized plasmas around TMR}\label{ss:MR:II}
 As presented in the last paragraph in the above section for Case~A and Case~C, the ionization rate is always higher than the recombination rate in Case~A, B, C, D and E. In this subsection, we will present the numerical results in Case~A0, in which have set the recombination rate to equal the ionization rate. By comparing Cases~A and A0, we demonstrate how the non-equilibrium ionization-recombination effect impacts magnetic reconnection in initially weakly ionized plasmas around the solar TMR. 
 
 Fig.~1(c) presents the distributions of the current density $J_z$ and ionization degree $f_i$ at $t=5.441$, $t=13.13$ and $t=23.381$ in Case~A0. One can see that the maximum current density during the later stage is higher than that in Case~A. However, the maximum ionization degree in Case~A0 can only reach around $0.08\%$, which is much lower than that in Case~A. The ionization rate increases extremely slowly with time in Case~A0. Fig.~2(c) shows the profiles of ion and neutral temperatures across the current sheet in Cases~A0 at the same three time instances as in Fig.~1(c). Both the plasma and neural temperatures reach very high values inside the current sheet. The maximum plasma temperature is above $5\times10^4$~K during the later stage of the reconnection process, which is much higher than that in Case~A. These results are consistent with the previous one fluid MHD simulations \citep{2016ApJ...832..195N}, where the plasmas were heated from $4200$~K to very high temperatures (about $8\times 10^4$~K) with reconnection magnetic fields of $500$~G at around the TMR region in that work. However, it is not realistic to have such high temperatures and weakly ionized hydrogen plasmas, which is a reflection of the unphysical nature of the ionization-recombination balance imposed in this calculation. Fig.~4(c) presents the total generated thermal energy and radiated energy inside the simulation domain in Case~A0. Joule heating is also the dominate term to generate the thermal energy in this case, and most of the generated thermal energy is also radiated away.  
 
The black dashed line in Fig. 6(a) represents the time dependent reconnection rate in Case~A0. The method for calculating the reconnection rate is the same as that presented in the above section. Though the maximum current density in Case~A0 is higher than that in Case~A, but the magnetic reconnection rate in Case~A0 is more than two times lower than that in Case~A. The much higher plasma temperature at the reconnection X-point in Case~A0 results in a much lower magnetic diffusion $\eta^{\ast}$, which is the main reason for a lower reconnection rate. 

By comparing the numerical results in Case~A and A0, one can conclude that the non-equilibrium ionization-recombination factor leads to much lower temperature increases inside the reconnection region. This factor also leads to a faster reconnection rate in our low Lundquist number simulations. Therefore, it is very important  and necessary to include the non-equilibrium ionization-recombination for studying the reconnection events around the solar TMR region, especially for answering the questions about how high the plasma temperature can be increased and if the plasma temperatures inside the reconnection region are high enough to produce Si IV emissions from the IRIS bombs around the solar TMR region. 
 
\subsection{The Hall effect on magnetic reconnection in initially weakly ionized plasmas around TMR }\label{ss:MR:III}

We next describe the roles of the Hall effect and the magnetic diffusion contributed by the electron-neutral collisions. We include the Hall effect and electron-neutral collisions in Case~B, and we performed calculations in the whole domain. We can then compare the evolution of each variable in the reconnection process in Cases~A and B. Fig.~8 shows the current density $J_z$, the ionization degree $f_i$ and the magnetic field in z-direction $B_z$ in the whole domain at three different times in Case~B. Comparing Figs.~1(a-c) and 8(a-f), we find that current density distributions in Cases~A and B are very similar and the maximum current density in Case~B is slightly higher than in Case~A. The maximum ionization degree can also reach about $40\%$ in Case~B. The electron-neutral collision can quickly heat the plasma temperature to above $4\times10^4$~K at the beginning, but then the plasma temperature drops fast to low values below $2\times10^4$~K. As shown in Fig.~9, the maximum temperature in Case~B eventually drops to around $1.5\times10^4$~K. Figs.~9(b) and 9(c) show the ion inflow velocity $V_{iy}$ and the difference in speeds of between ion and neutral inflow $V_{iy}-V_{ny}$ at $x=0$ along $y$-direction at three different times. Comparing Figs.~5(a) and 5(b) with Figs.~9(b) and 9(c), we notice that the plasma inflow velocity distributions are also very similar in Cases~A and B, the maximum of $V_{iy}$ in Case~B is slightly higher than that in Case~A. The plasma and the neutral inflows are also coupled well in the reconnection upstream region in Case~B. 

As shown in Fig.~8, the Hall effect does not result in the obvious asymmetry in either $x$ or $y$ directions for any of  the variables. Tilting of the current sheet can only occur on the scales of the ion inertial length $d_i$, and $d_i$ decreases to a very small value, even smaller than the grid size during the reconnection process in our simulations. Therefore,  a tilted current sheet as shown by \cite{2003PhPl...10.3131L} does not appear in Case~B in this work. From Fig.~6(a), we can find that the Hall effect just slightly enhances the reconnection rate. Fig.~4(b) shows that the electron-neutral collision only strongly affects the magnetic diffusion at the very beginning and the magnetic diffusion coefficient contributed by electron-neutral collisions quickly drops to a low value before the reconnection process starts. Values of $\eta^{\ast}$ are eventually very close to one another in Cases~A and B. Though the Hall effect increases the reconnection rate slightly and it enhances the inflow velocity and the maximum current density in the same fashion correspondingly more apparent in Case~B than in Case~A, including the Hall effect and electron-neutral collisions do not significantly change the temporal evolution and the spatial structure of the reconnection process. 

\section{Summary and discussions}\label{s:discussion}
Magnetic reconnection in strongly magnetized regions around the solar TMR has been studied by using the reactive multi-fluid plasma-neutral module of the HiFi modeling framework. Several cases with different strengths of magnetic field have been simulated. We have examined the impact of non-equilibrium ionization-recombination physics by comparing two cases: one that allows for appropriate reactive dynamical evolution, and another where the recombination rate was set to be equal to the ionization rate. We have also compared two simulations with the same strength of magnetic field, but with the Hall term and electron-neutral collisions included in only one of them. From the numerical results we summarize four main conclusions as follows:

(1) In initially weakly ionized plasmas around the solar TMR region, it is necessary to include the non-equilibrium ionization-recombination for studying and comparing with observations of the reconnection events around the solar TMR region. The non-equilibrium ionization-recombination effect leads to much lower temperature increases inside the reconnection region. This effect also leads to a faster reconnection rate before plasmoid instabilities appear.

 (2)  In a low $\beta$ plasma and in the absence of magnetic nulls, the ionized and neutral fluid flows are well-coupled throughout the reconnection region. They begin to decouple in the reconnection upstream region  when the initial plasma $\beta$ is high enough ($\beta_0=1.46$), and the decoupling of ionized and neutral fluid flows make the reconnection rate much faster than those in the low $\beta$ cases.
 
 (3) In the absence of magnetic nulls, the rate of ionization of the neutral component of the plasma is faster than recombination within the current sheet region even when the initial plasma $\beta$ is as high as $\beta_0=1.46$. When the reconnecting magnetic field is in excess of a kilogauss and the plasma $\beta$ is lower than 0.0145, the initially weakly ionized plasmas can become fully ionized and the current sheet can be strongly heated to above $2.5\times10^4$~K within the reconnection region, even as most of the collisionally dissipated magnetic energy is radiated away.
 
 (4) The Hall effect increases the reconnection rate slightly, but it does not result in significant asymmetries or change the characteristics  of the current sheet in the reconnection process.   

As more and more small scale reconnection events are observed through the high resolution solar telescopes, magnetic reconnection in the low solar atmosphere with partially ionized plasma is becoming very important for understanding how the solar activity is generated. More realistic reconnection models are essential for answering the question of whether or not it is possible for magnetic reconnection process near the temperature minimum region to heat plasma to a high enough temperature to produce the emission identified in Si IV slit-jaw observations. The bright points  and small scale jets which are frequently observed in sunspot regions prove that magnetic field strengths can be as high as several kilogauss during reconnection in the low solar atmosphere. It is therefore very likely that low $\beta$ reconnection as presented in this paper can frequently happen near the temperature minimum region. Our results show that the plasma and neutral fluids in the low $\beta$ plasma can be well coupled in both the inflow and outflow regions of the reconnection process. The current sheet is almost in steady state, and ionization always dominates over recombination inside the current sheet region even when the plasma $\beta$ is as high as $\beta=1.46$. The plasma can be fully ionized as long as the plasma $\beta$ is low enough. The plasma inside the current can then be heated from several thousand kelvin to above $4\times10^4$~K. Under these conditions, the highly ionized plasma inside the current sheet makes the electron-neutral collisions less important and the Hall effect does not appear to be significant in the reconnection process. 

One should note that magnetic reconnection in EBs or IRIS bombs may also happen in the photosphere, where the plasma density could be 10-50 times higher than that in our simulations around the TMR region. The much stronger radiative cooling in the photosphere will make it much more difficult to heat the plasma in magnetic reconnection process. In order to compare with the observations, larger-scale simulations are necessary for the future studies. In addition to the ground state and the ionized state, the excited state of the hydrogen gas should also be included. Background heating is not included in this model, and the line radiative cooling term in this work radiated most of the generated thermal energy, which makes the plasma difficult to heat. A more realistic representation of radiative cooling will be important for understanding magnetic reconnection in the low solar atmosphere. 
 
\section*{ACKNOWLEDGMENTS}
The authors are grateful to the referee for the valuable comments and suggestions. Lei Ni would like to thank Dr. Hui Tian for his helpful discussions. This research is supported by NSFC Grants 11573064, 11203069, 11333007, 11303101 and 11403100; the Western Light of Chinese Academy of Sciences 2014; the Youth Innovation Promotion Association CAS 2017; Program 973 grant 2013CBA01503; NSFC-CAS Joint Grant U1631130; CAS grant QYZDJ-SSW-SLH012; and the Special Program for Applied Research on Super Computation of the NSFC-Guangdong Joint Fund (nsfc2015-460, nsfc2015-463, the third phase). The authors gratefully acknowledge the computing time granted by the Yunnan Astronomical Observatories and the National Supercomputer Center in Guangzhou, and provided on the facilities at the Supercomputing Platform, as well as the help from all faculties of the Platform. J.L. also thanks the help from the National Supercomputing Center in Tianjin. V.S.L. acknowledges support from the US National Science Foundation (NSF). N.A.M. acknowledges support from NSF SHINE grant AGS-135842 and DOE grant DE-SC0016363.

\clearpage

\begin{table}
 \caption{The differences in initial conditions and evolution equations among the six simulation cases.}
  \label{Parameter}
   \begin{tabular}{lcccccc}
       \hline
                            &  Case~A & Case~A0 &   Case~B    &    Case~C    &   Case~D  &  Case~E          \\
       \hline
          $b_p$        &   1              &       1        &         1           &       0.2          &       2           &      3                 \\ 
       \hline  
          with Hall and $\nu_{en}$ &  No & No & Yes & No & No & No        \\
        \hline
           with non-equilibrium \\
           ionization and \\
           recombination      &  Yes & No & Yes & Yes & Yes & Yes        \\
       \hline
   \end{tabular}
\end{table}

  \begin{figure}
      \centerline{\includegraphics[width=0.33\textwidth, clip=]{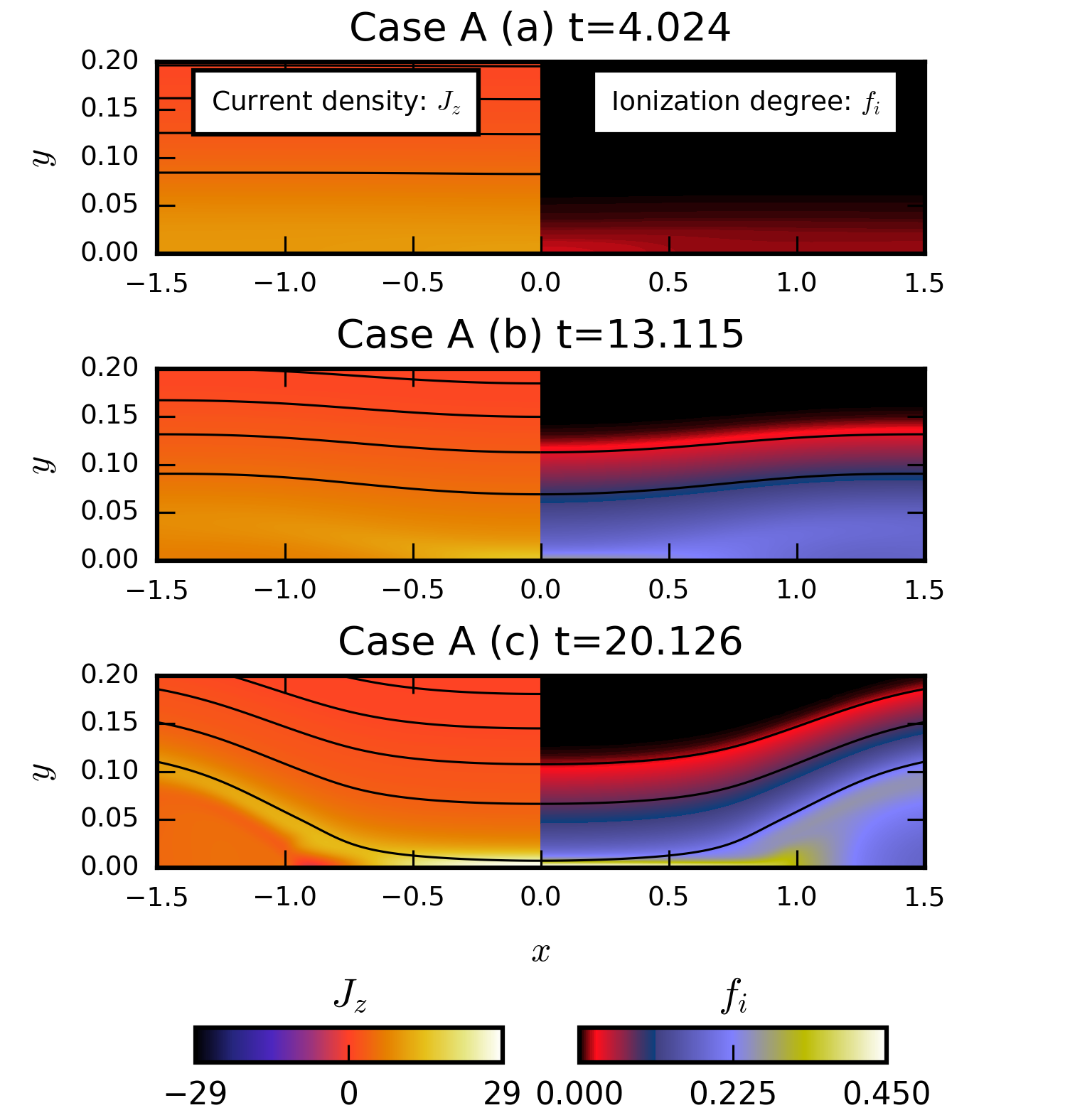}
                          \includegraphics[width=0.33\textwidth, clip=]{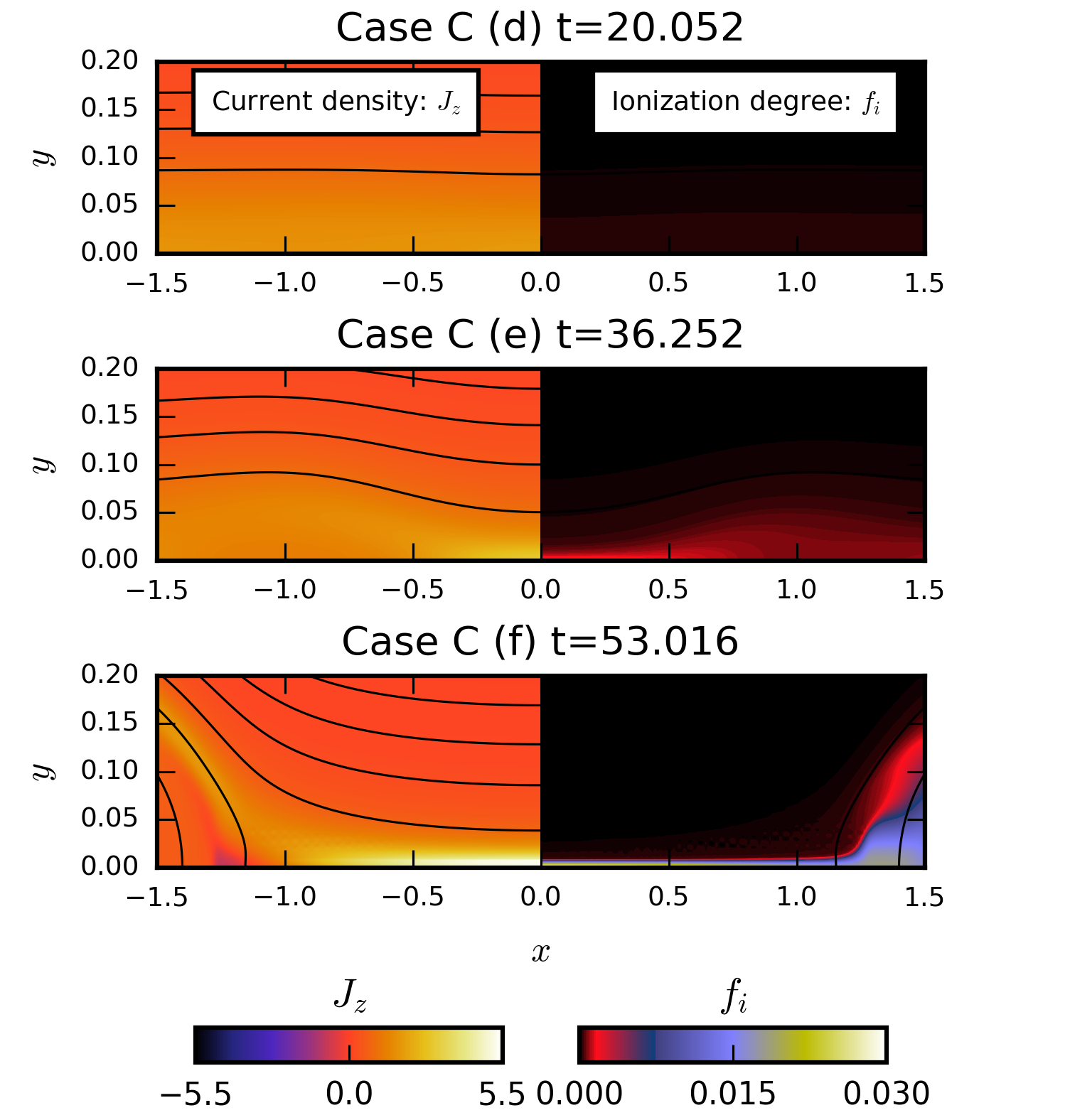}
                            \includegraphics[width=0.33\textwidth, clip=]{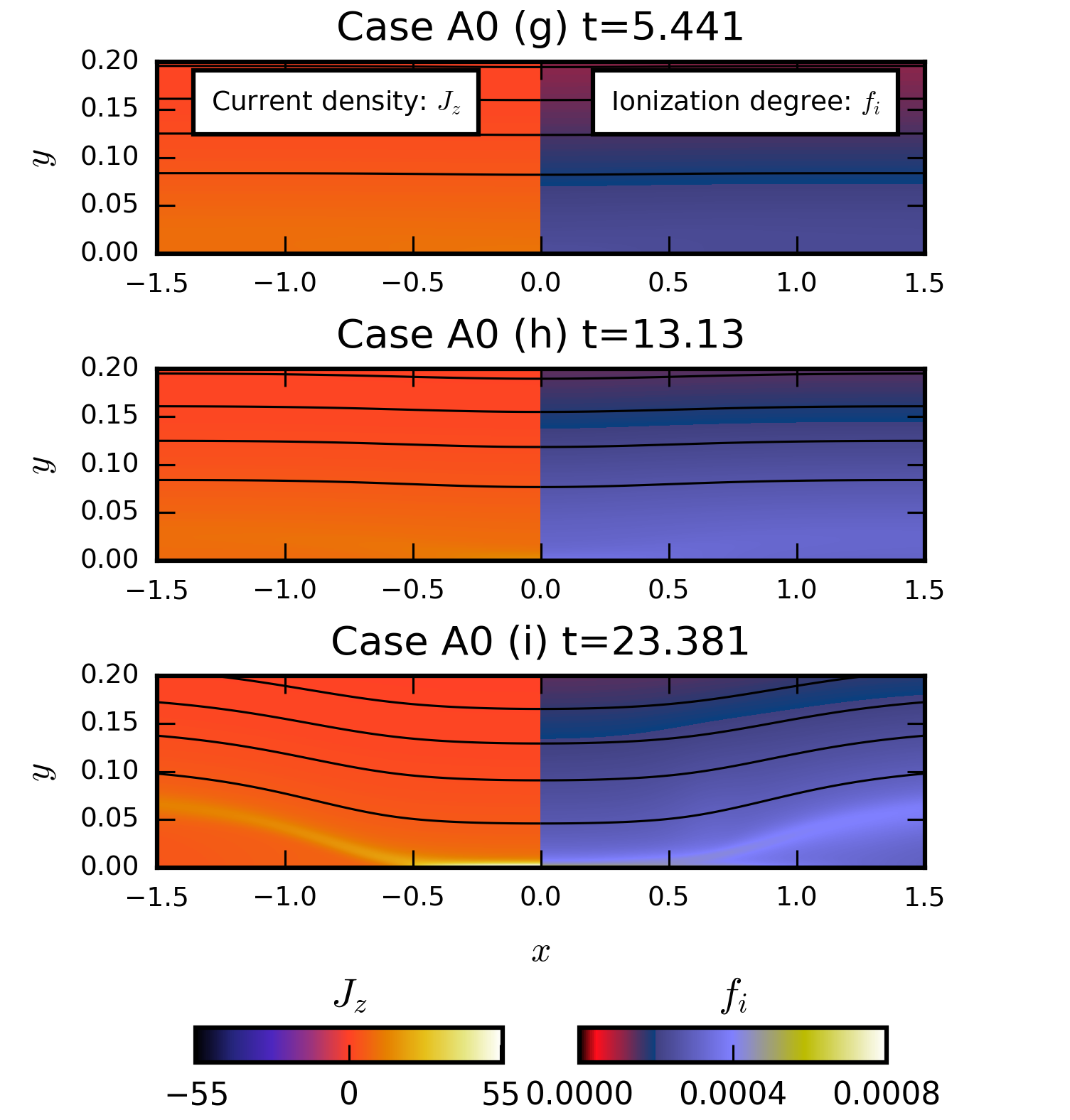} }
      \caption{(a), (b) and (c) show the current density $J_z$ (left) and ionization degree $f_i$ (right)  in one quarter of the domain at $t=4.024$, $t=13.115$ and $t=20.126$ in Case~A; (d), (e) and (f) show the same at $t=20.052$, $t=36.252$ and $t=53.016$ in Case~C; (g), (h) and (i) show the same at $t=5.441$, $t=13.13$ and $t=23.381$ in Case~A0. The black contour lines represent the out of plane component of the magnetic flux $A_z$ in these 2D figures. }
   \label{fig.1}
\end{figure}
  
 \begin{figure}
      \centerline{\includegraphics[width=0.33\textwidth, clip=]{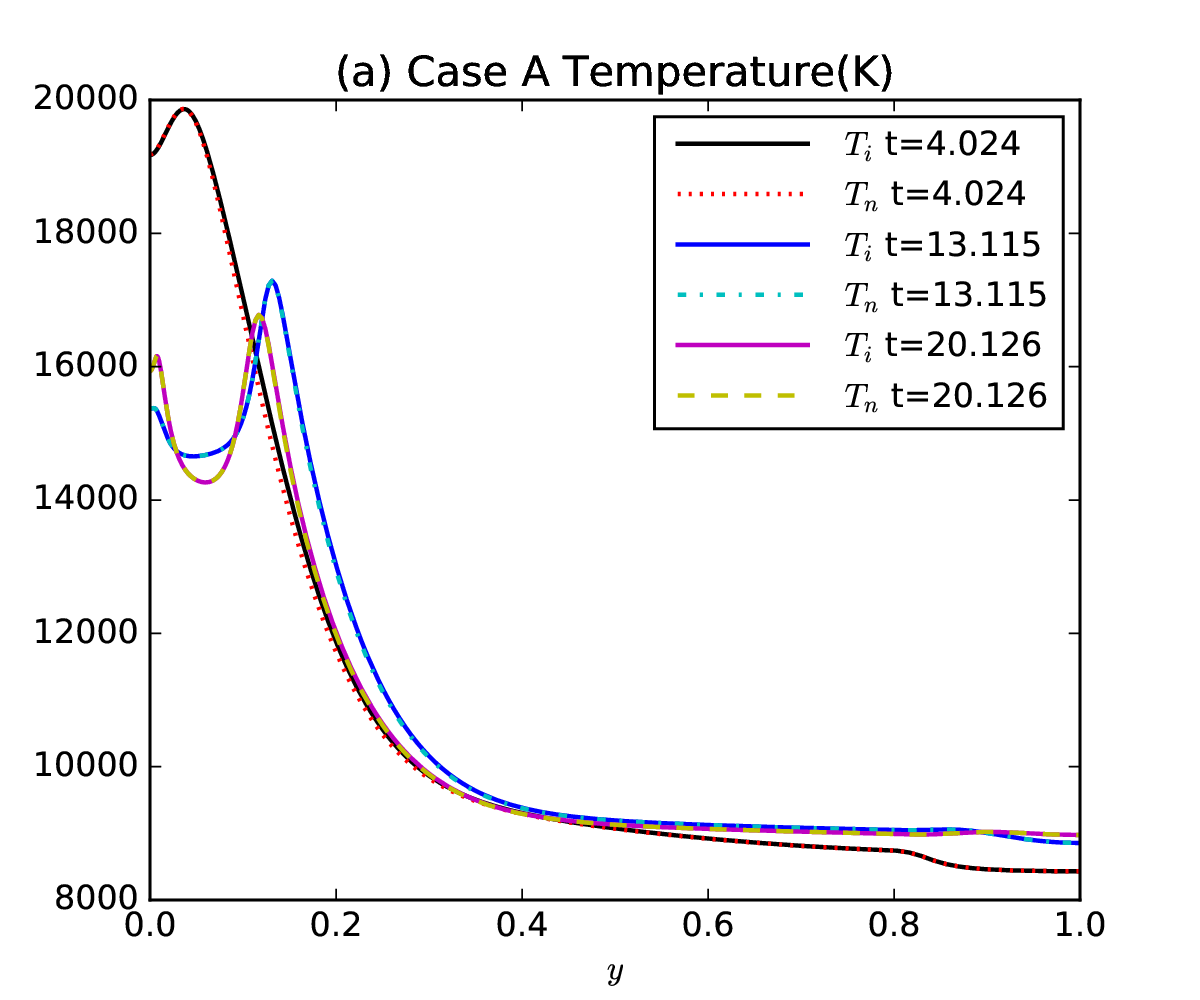}
                          \includegraphics[width=0.33\textwidth, clip=]{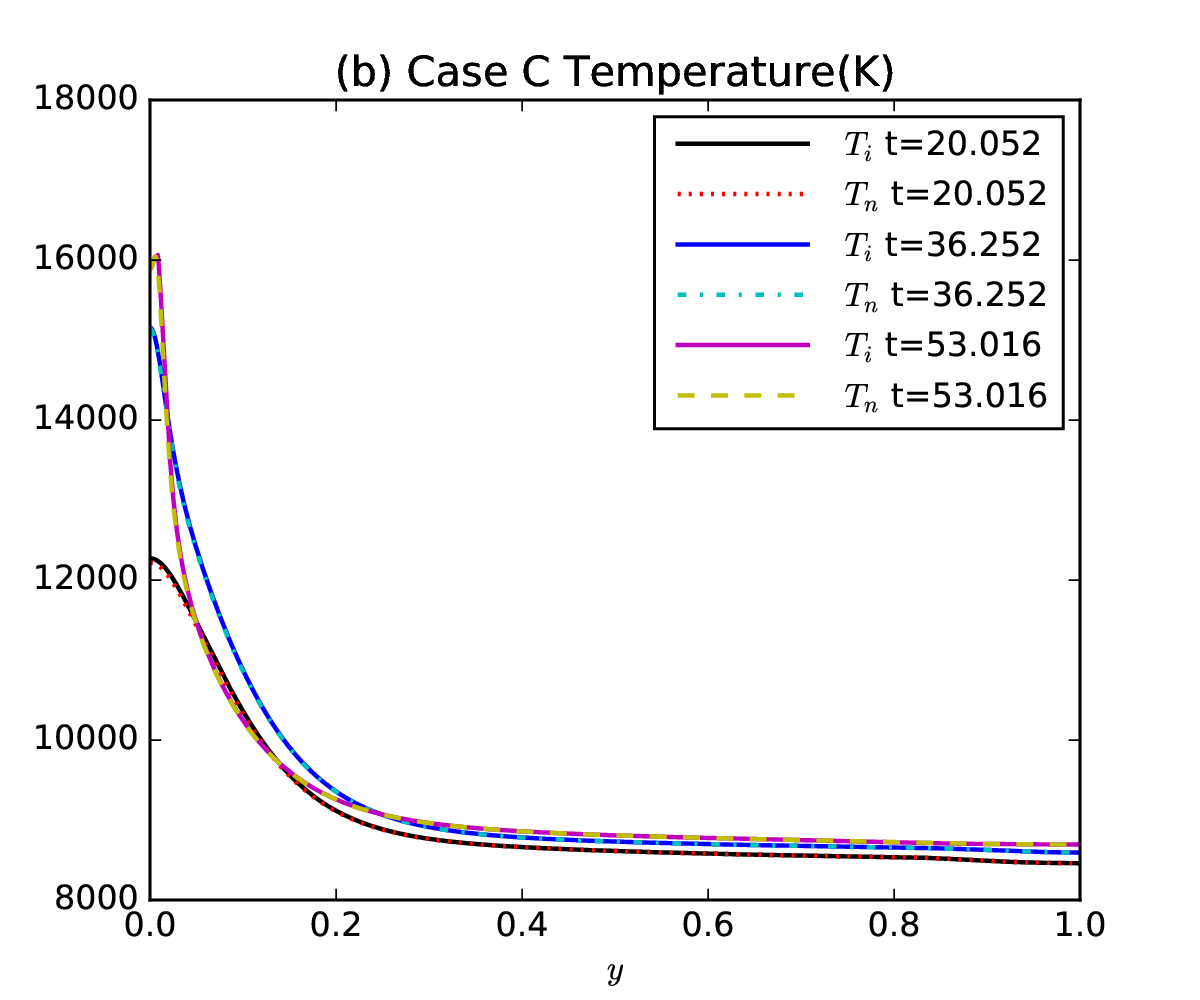} 
                          \includegraphics[width=0.33\textwidth, clip=]{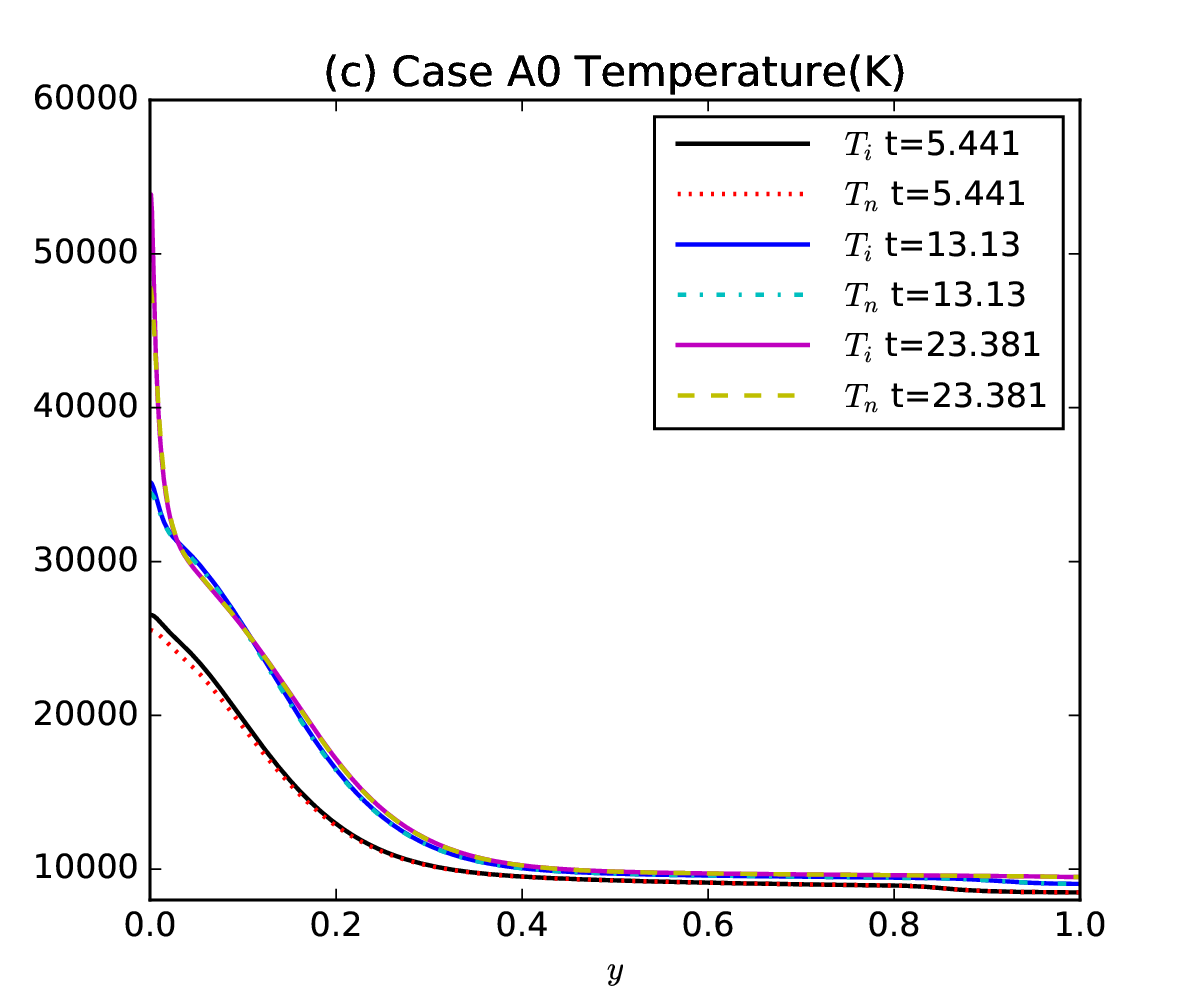}}
      \caption{(a) shows the distributions of the ion temperature $T_i$ and neutral temperature $T_n$ in Kelvin at $x=0$ along y direction at $t=4.024$, $t=13.115$ and $t=20.126$ in Case~A; (b) shows the same at $x=0$ along y direction at $t=20.052$, $t=36.252$ and $t=53.016$ in Case~C; (c) shows the same at $x=0$ along y direction at $t=5.441$, $t=13.13$ and $t=23.381$ in Case~A0.}
    \label{fig.2}
\end{figure} 
   
\begin{figure}
      \centerline{\includegraphics[width=0.5\textwidth, clip=]{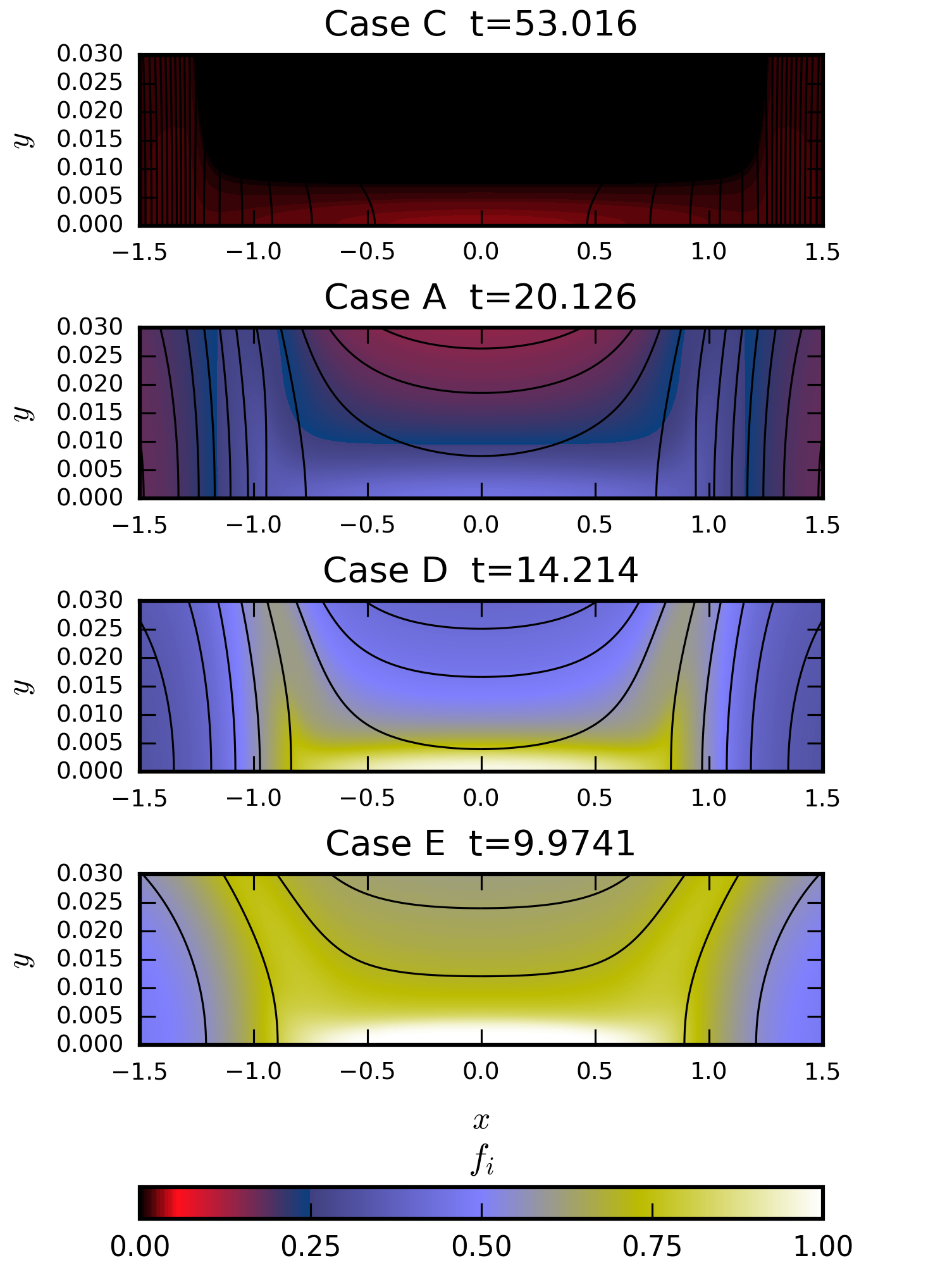}
                          \includegraphics[width=0.5\textwidth, clip=]{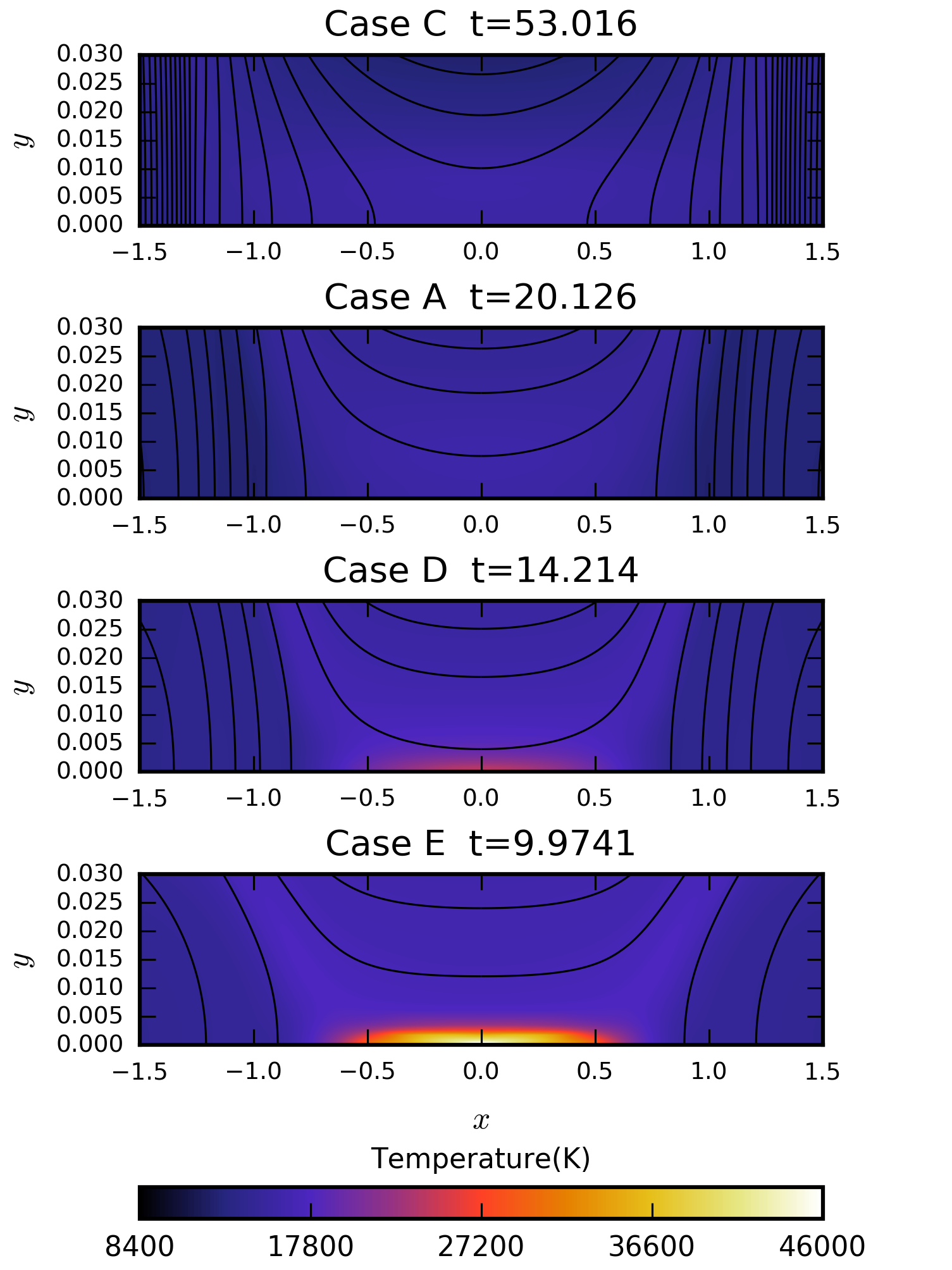} }
      \caption{The left panel and the right panel individually show the ionization degree $f_i$ and the neutral temperature $T_i$ inside the current sheet region in half of the domain in Case~C at $t=53.016$, in Case~A at $t=20.126$, in Case~D at $t=14.214$ and in Case~E at $t=9.9741$. }
     \label{fig.3}
\end{figure}   
  
\begin{figure}
      \centerline{\includegraphics[width=0.33\textwidth, clip=]{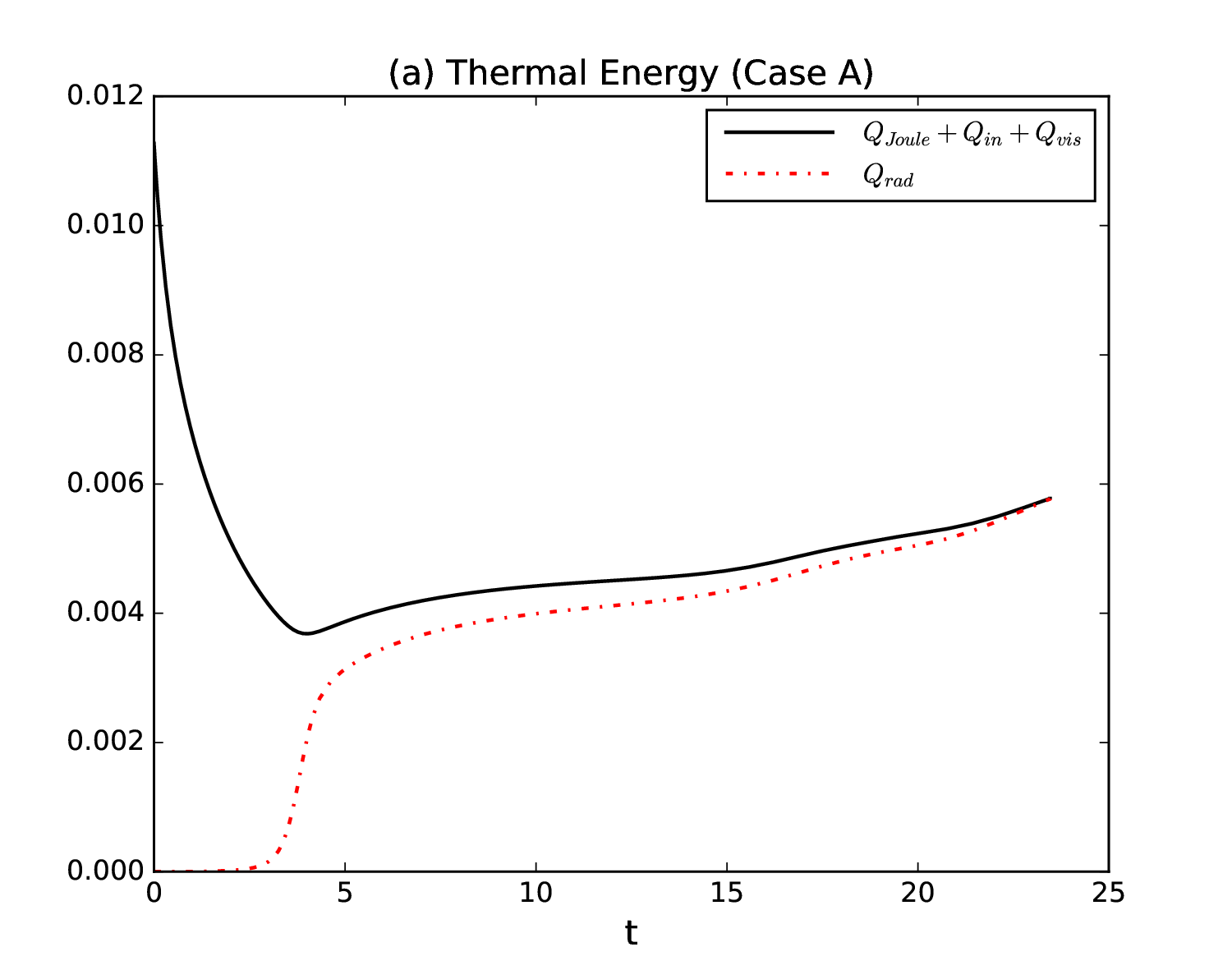}
                           \includegraphics[width=0.33\textwidth, clip=]{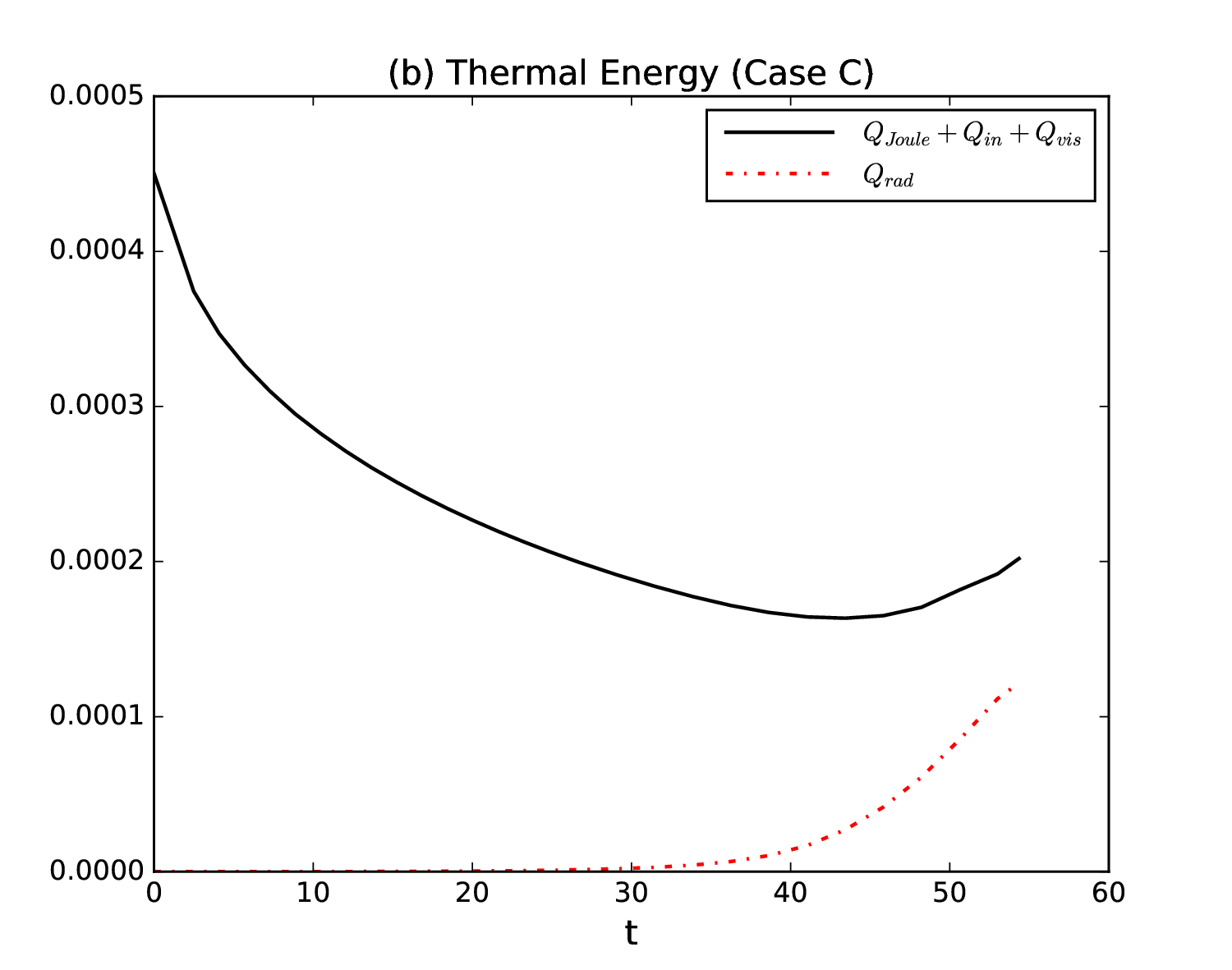}
                           \includegraphics[width=0.33\textwidth, clip=]{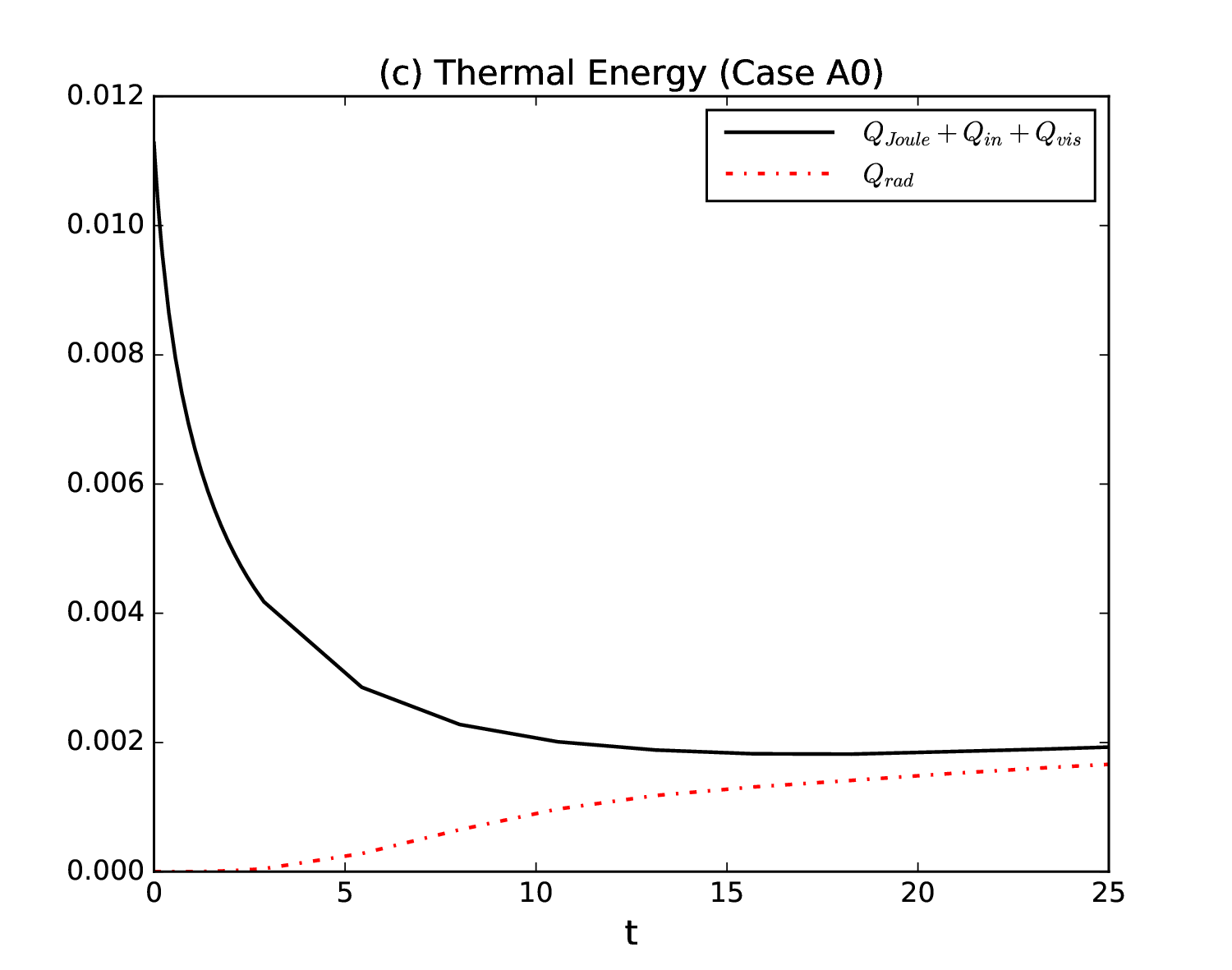}}
      \caption{(a) shows the time dependent total thermal energy and the energy loss as a result of radiation inside the reconnection domain in Case~A; (b) same as panel (a) for Case~C; (c) same as panel (a) for Case~A0. }
    \label{fig.4}
\end{figure}

 \begin{figure}
      \centerline{\includegraphics[width=0.3\textwidth, clip=]{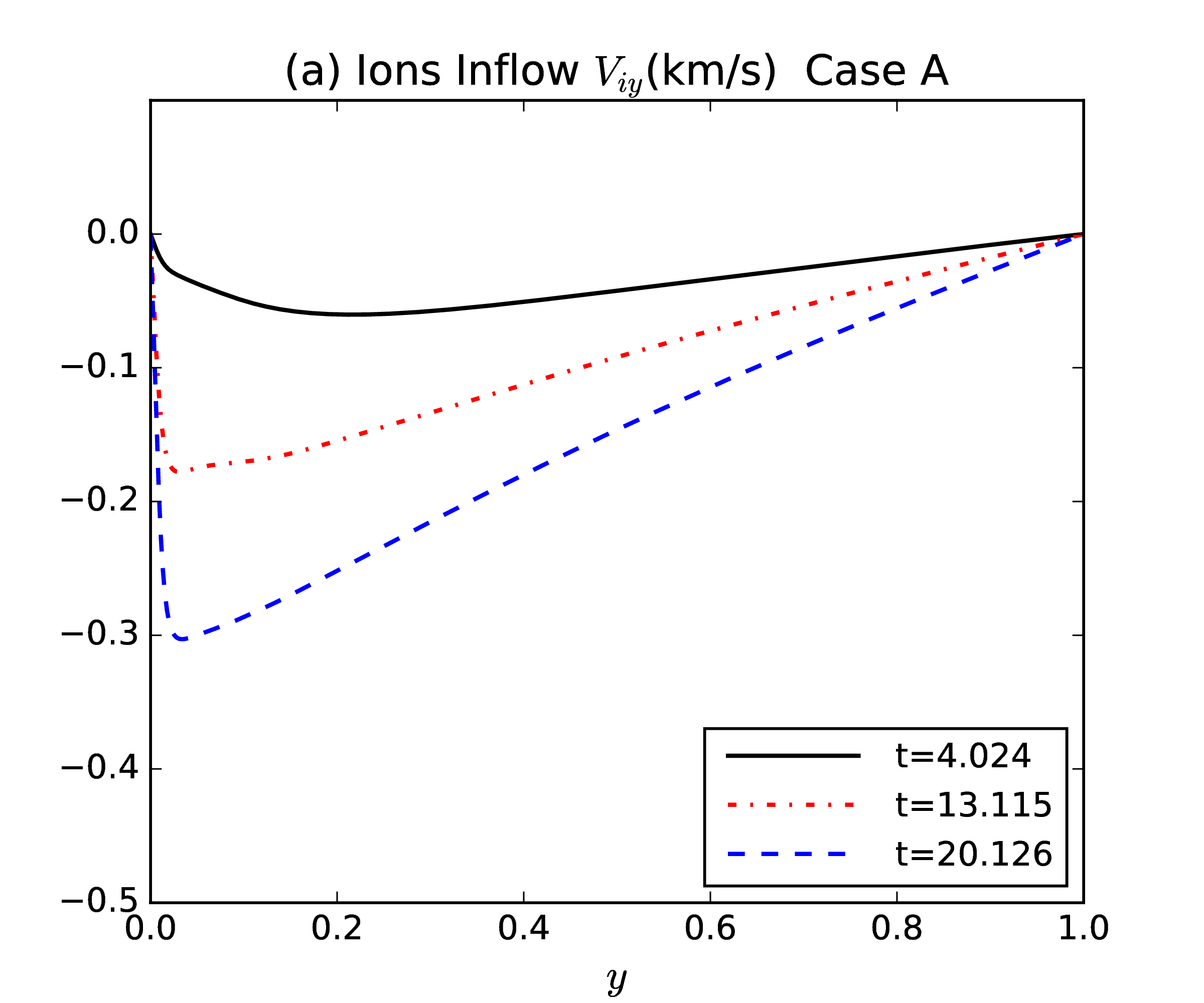}
                          \includegraphics[width=0.3\textwidth, clip=]{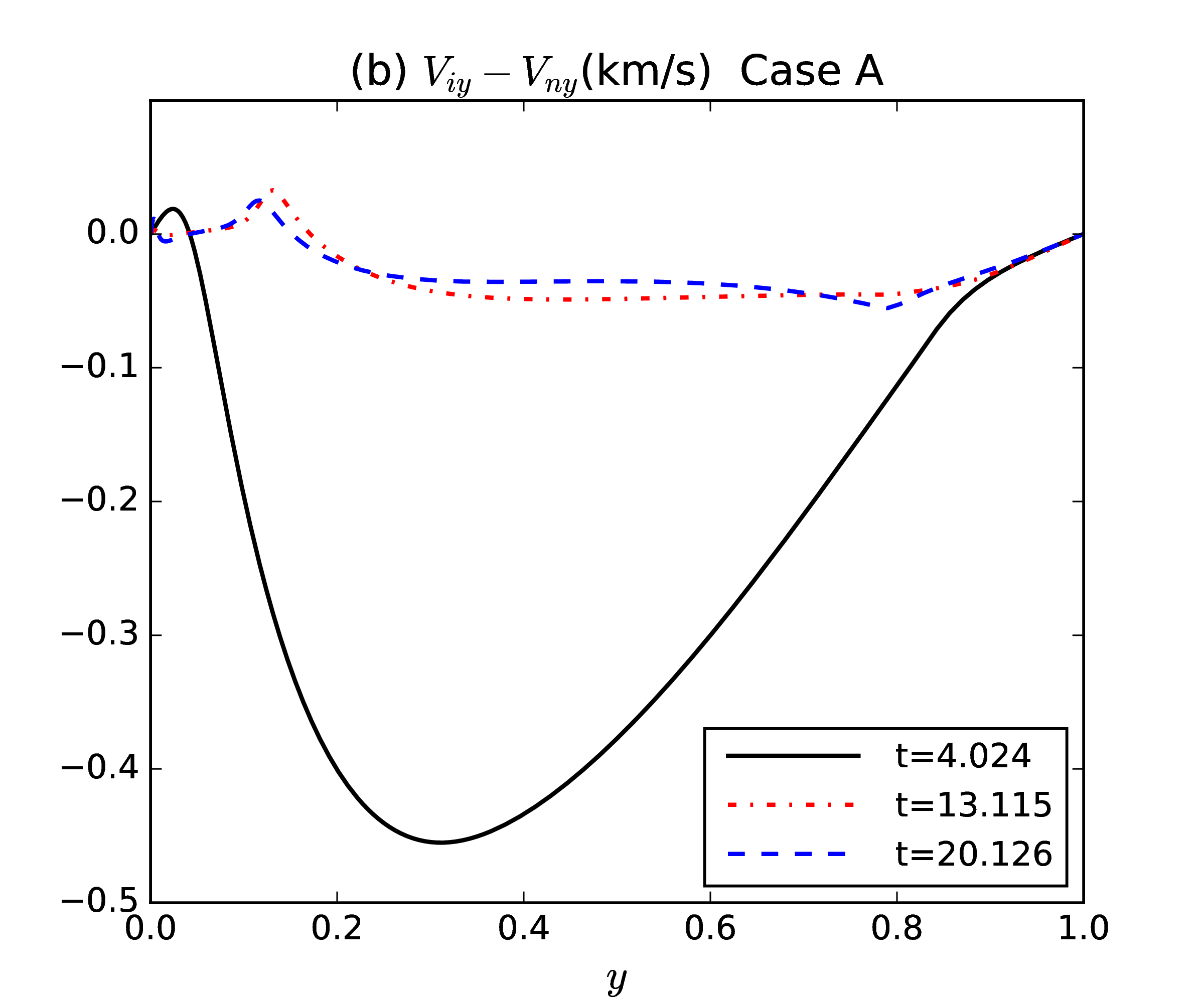} 
                           \includegraphics[width=0.3\textwidth, clip=]{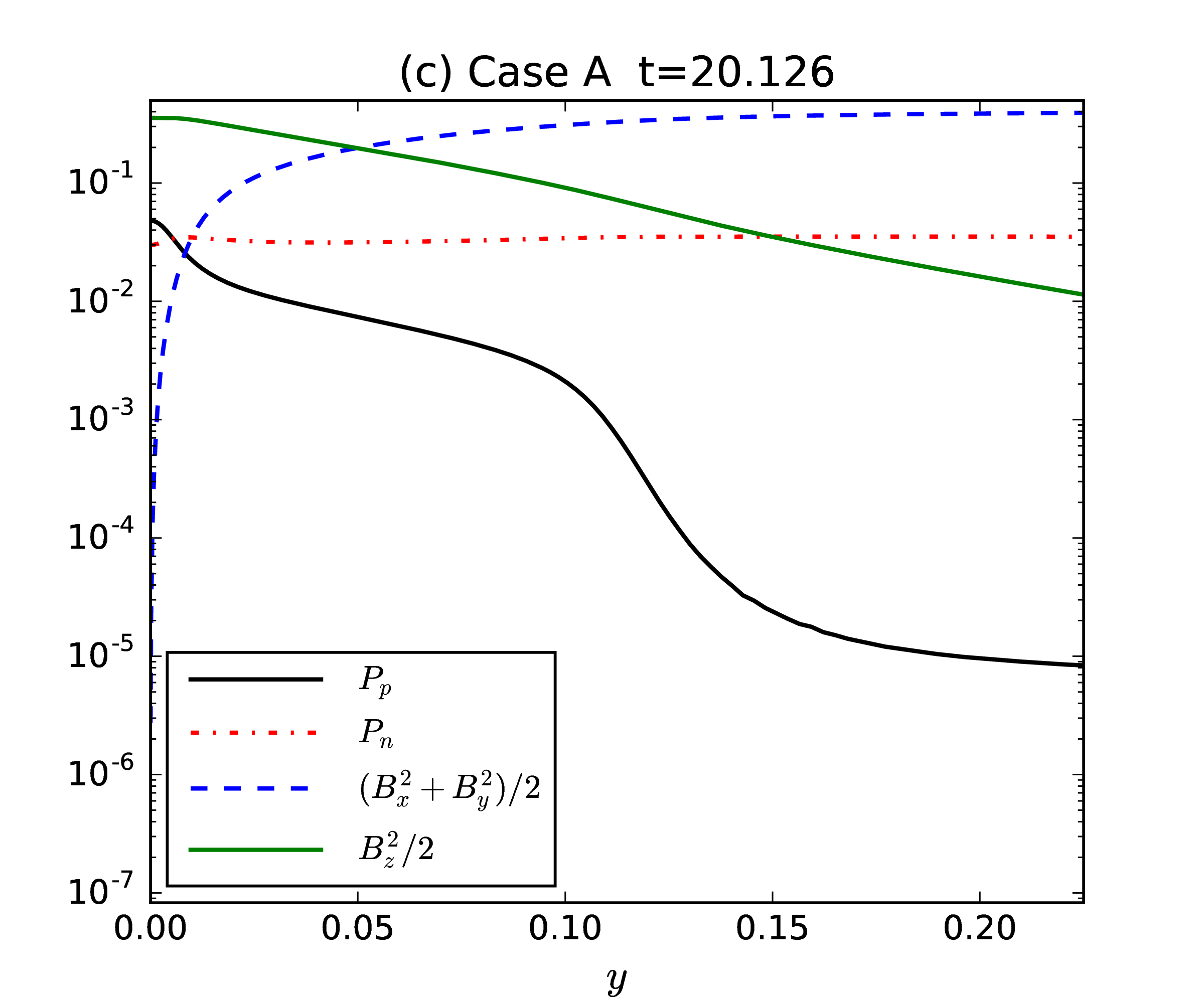} }
      \centerline{\includegraphics[width=0.3\textwidth, clip=]{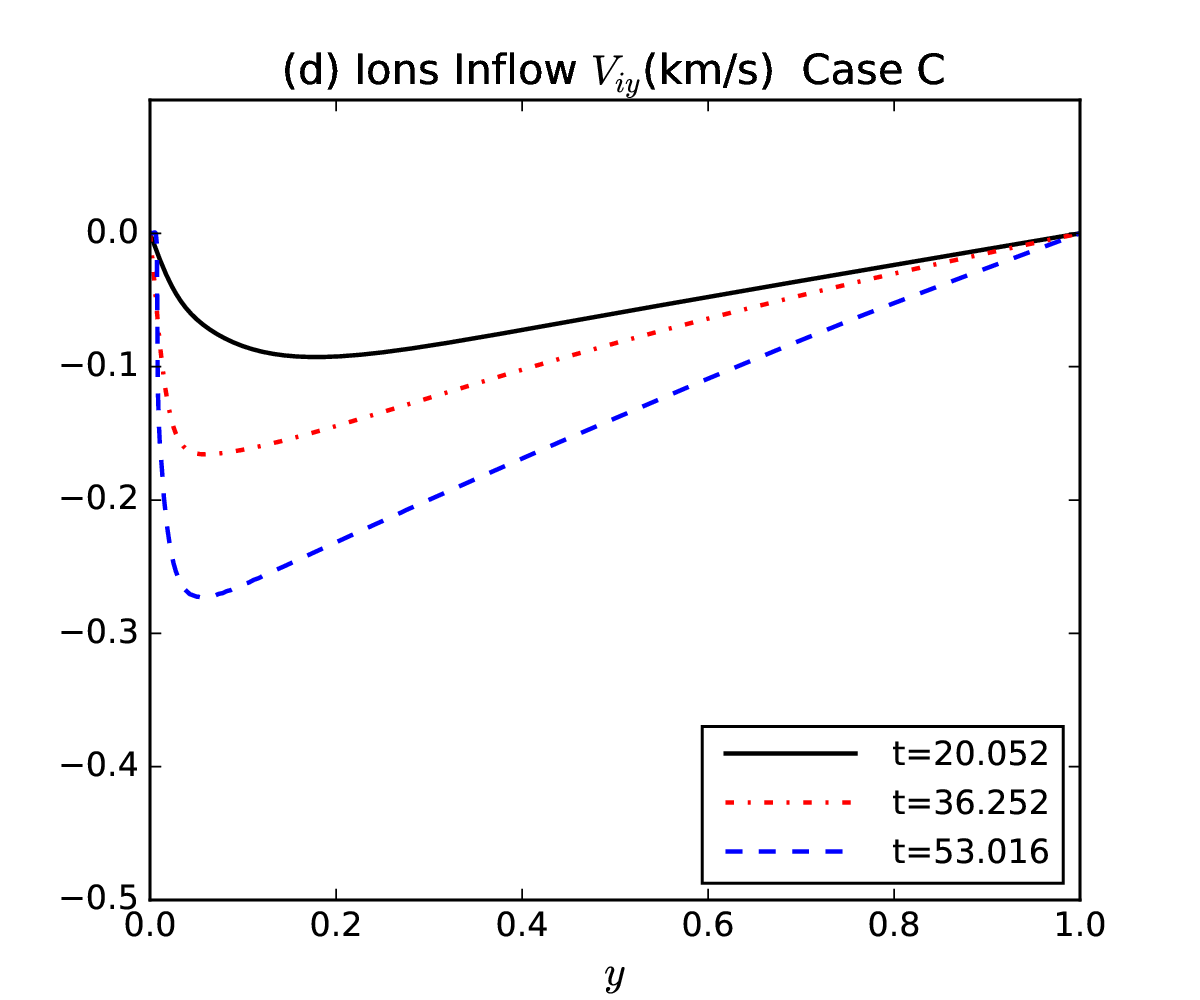}
                          \includegraphics[width=0.3\textwidth, clip=]{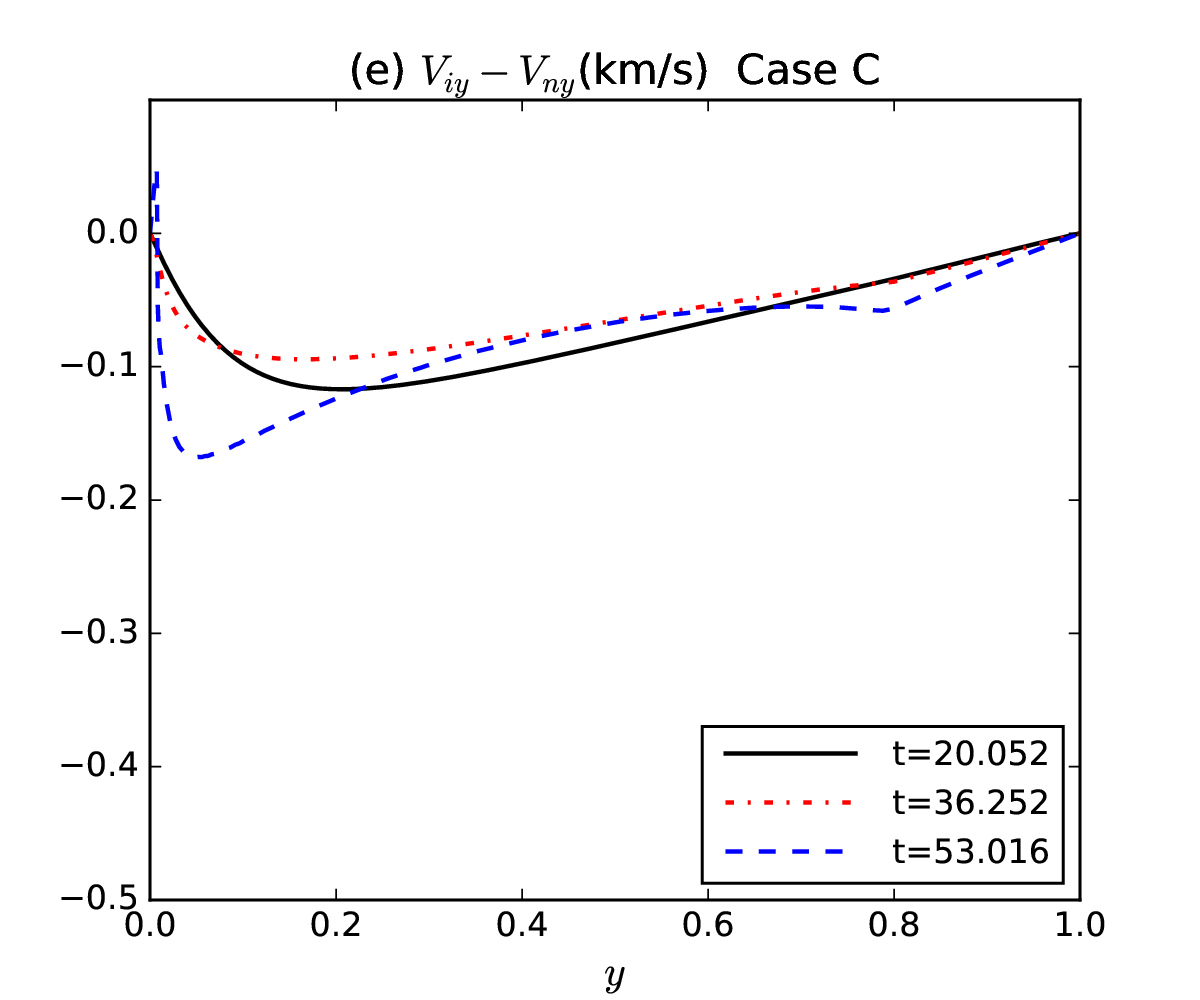} 
                           \includegraphics[width=0.3\textwidth, clip=]{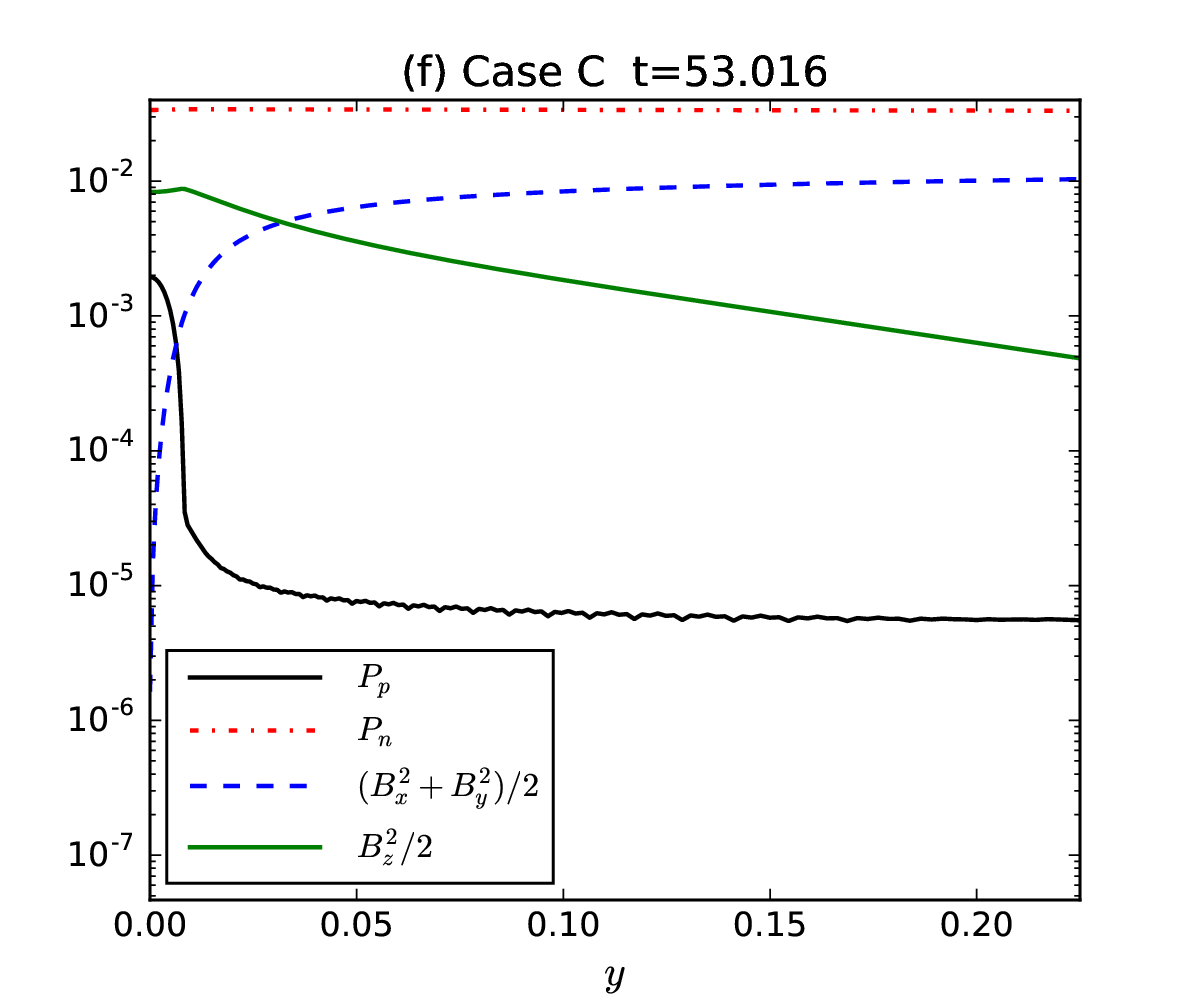} }
      \caption{Panels (a) and (b) show the ion inflow speed $V_{iy}$, and the difference in speed between the ion and the neutral inflows $V_{iy}-V_{ny}$ across the current sheet with dimensions at $x=0$  at $t=4.024$, $t=13.115$ and $t=20.126$ in Case~A. Panel (c) shows the terms contributing to the pressure balance across the current sheet at x=0 during the quasi-steady-state phase at $t=20.126$ in Case~A. Panels (d) and (e) similarly show  the ion inflow $V_{iy}$ and $V_{iy}-V_{ny}$ across the current sheet at $x=0$ at $t=20.052$, $t=36.252$ and $t=53.016$ in Case~C; and panel (f) shows the pressure balance across the quasi-steady-state current sheet at $t=53.016$ in Case~C. Note that the range of $y$ is from $0\leq y \leq 0.225$ in panels (c) and (f) to show clearly the contributions from different terms to the pressure balance within the current sheet. }
     \label{fig.5}
\end{figure} 

\begin{figure}
     \centerline{\includegraphics[width=0.33\textwidth, clip=]{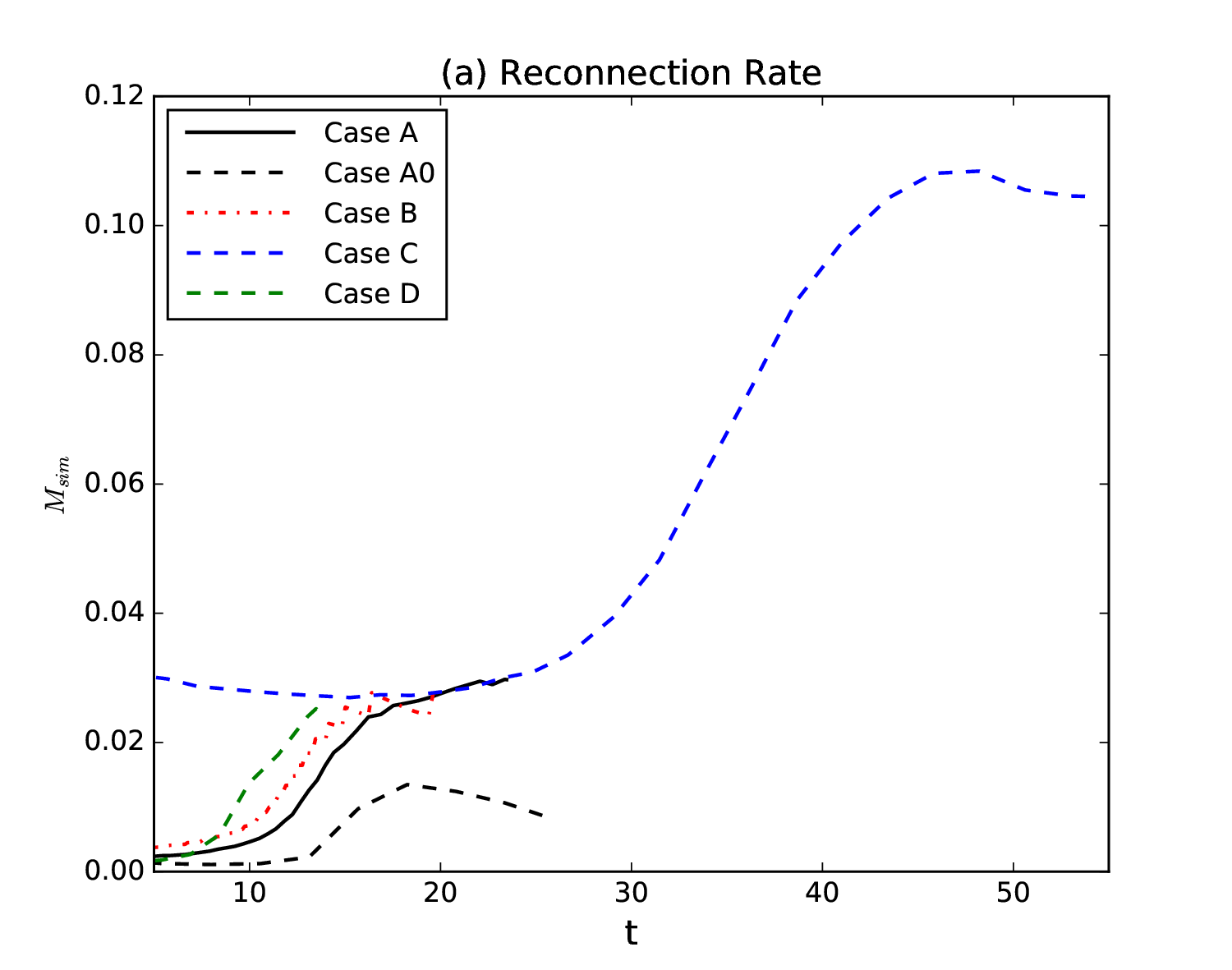}
                           \includegraphics[width=0.33\textwidth, clip=]{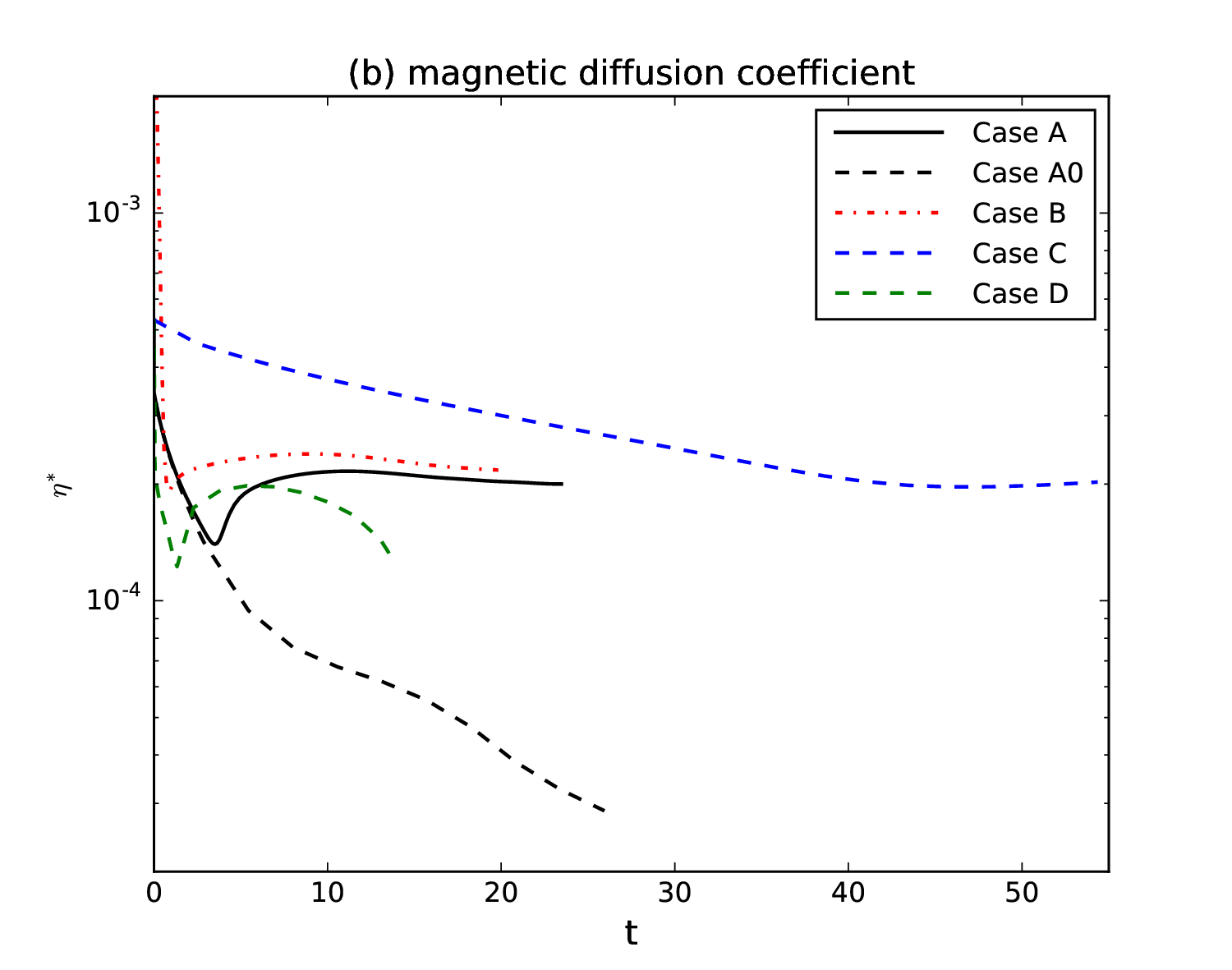} }
     \caption{(a) The time dependent reconnection rate in Cases~A, A0, B, C and D; (b) the time dependent magnetic diffusion coefficient $\eta^{\ast}$ at the reconnection X-point in Cases~A, A0, B, C and D. }
     \label{fig.6}
\end{figure}

 \begin{figure}
      \centerline{\includegraphics[width=0.33\textwidth, clip=]{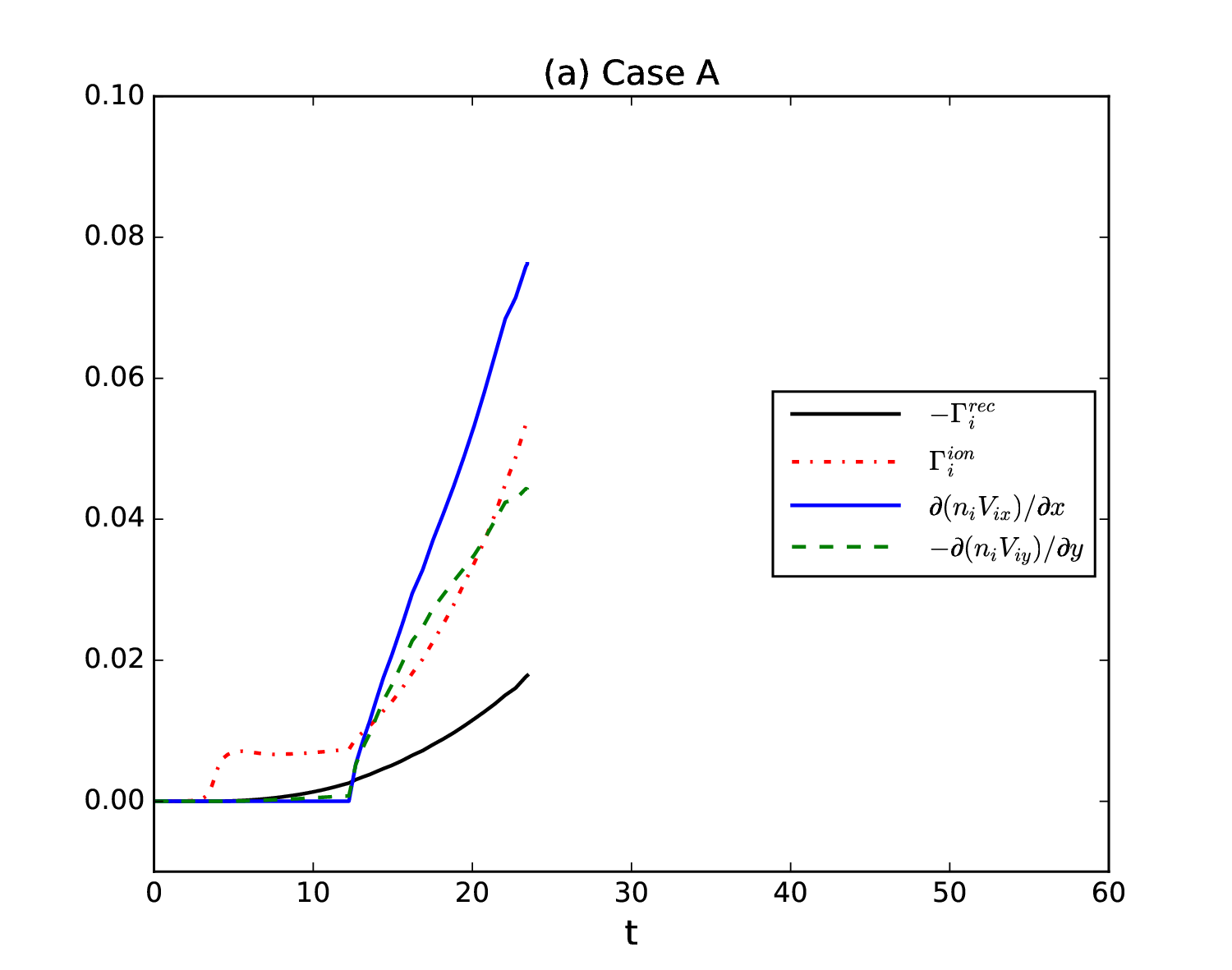}
                           \includegraphics[width=0.33\textwidth, clip=]{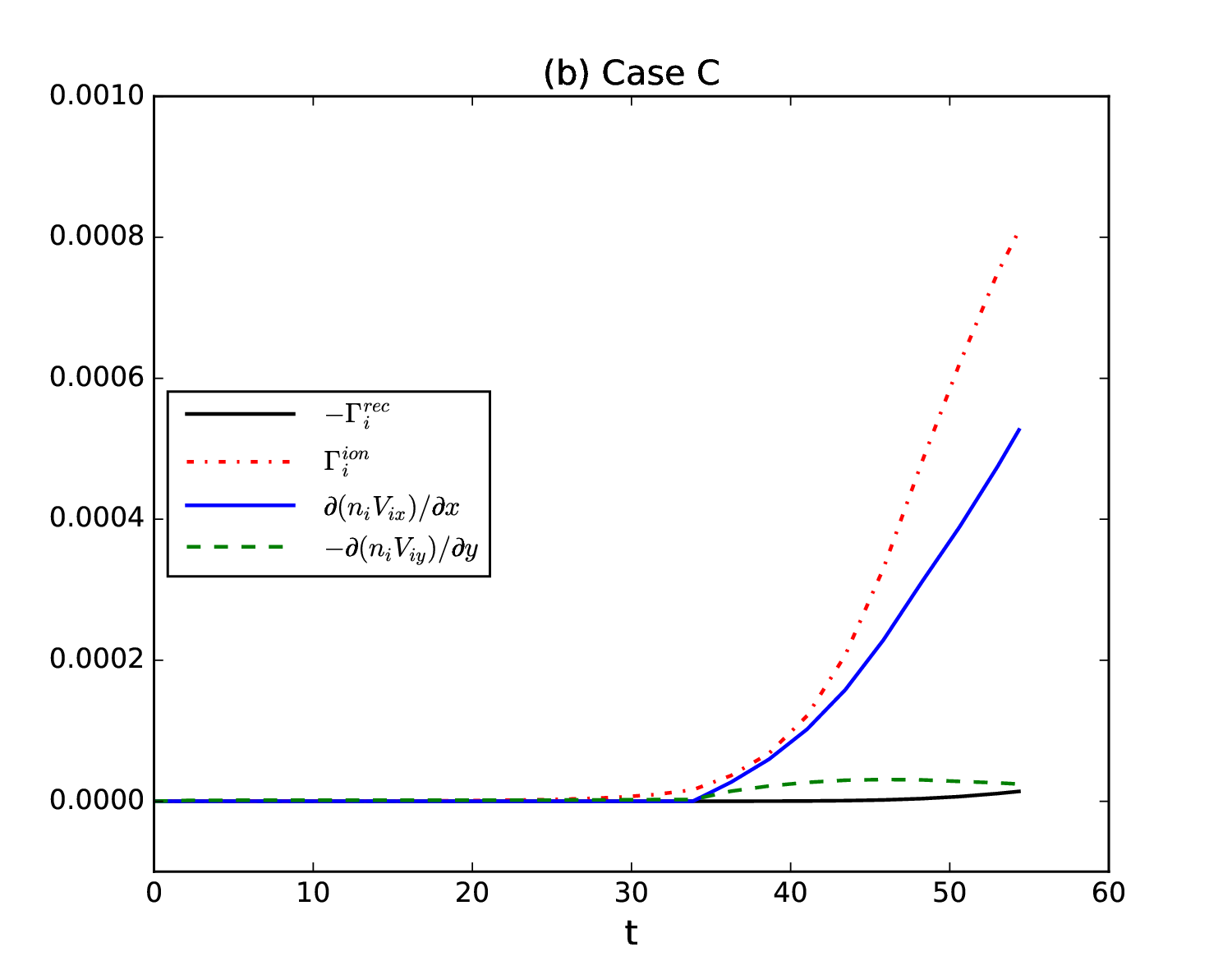}
                           \includegraphics[width=0.33\textwidth, clip=]{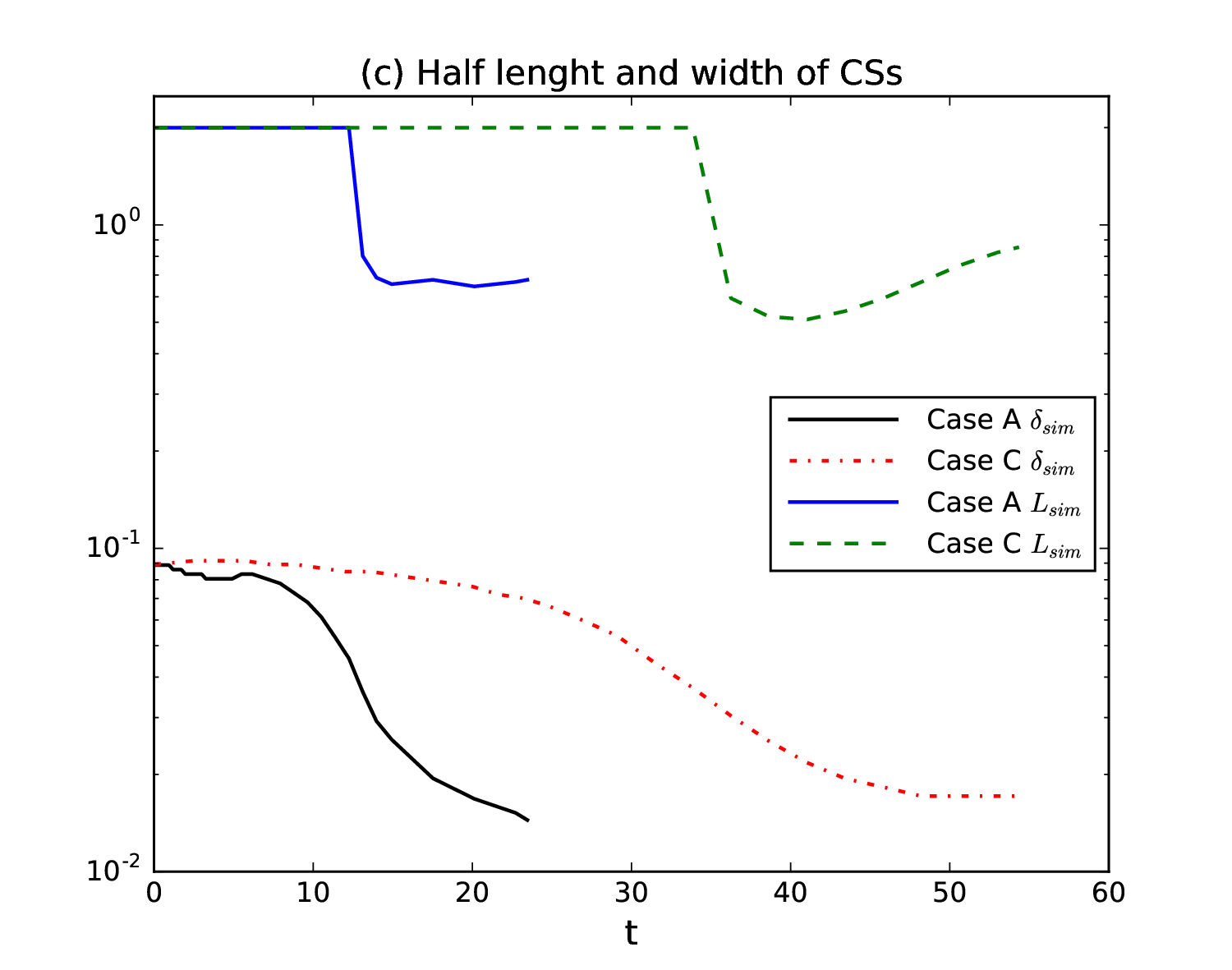} }
       \caption{(a) The same time dependent contributions to $\partial n_i/\partial t$ in Case~A; (b) the same in Case~C. The four contributions are the average values inside the current sheet, the loss due to recombination $-\Gamma_i^{rec}$;  the loss due to the outflow $\partial(n_iV_{ix})/\partial x$, the gain due to the inflow $-\partial(n_iV_{iy})/\partial y$, and the gain due to ionization $\Gamma_i^{ion}$. (c) The half length and width of CSs in Cases~A and C. }
    \label{fig.7}
\end{figure} 
  
 \begin{figure}
      \centerline{\includegraphics[width=0.33\textwidth, clip=]{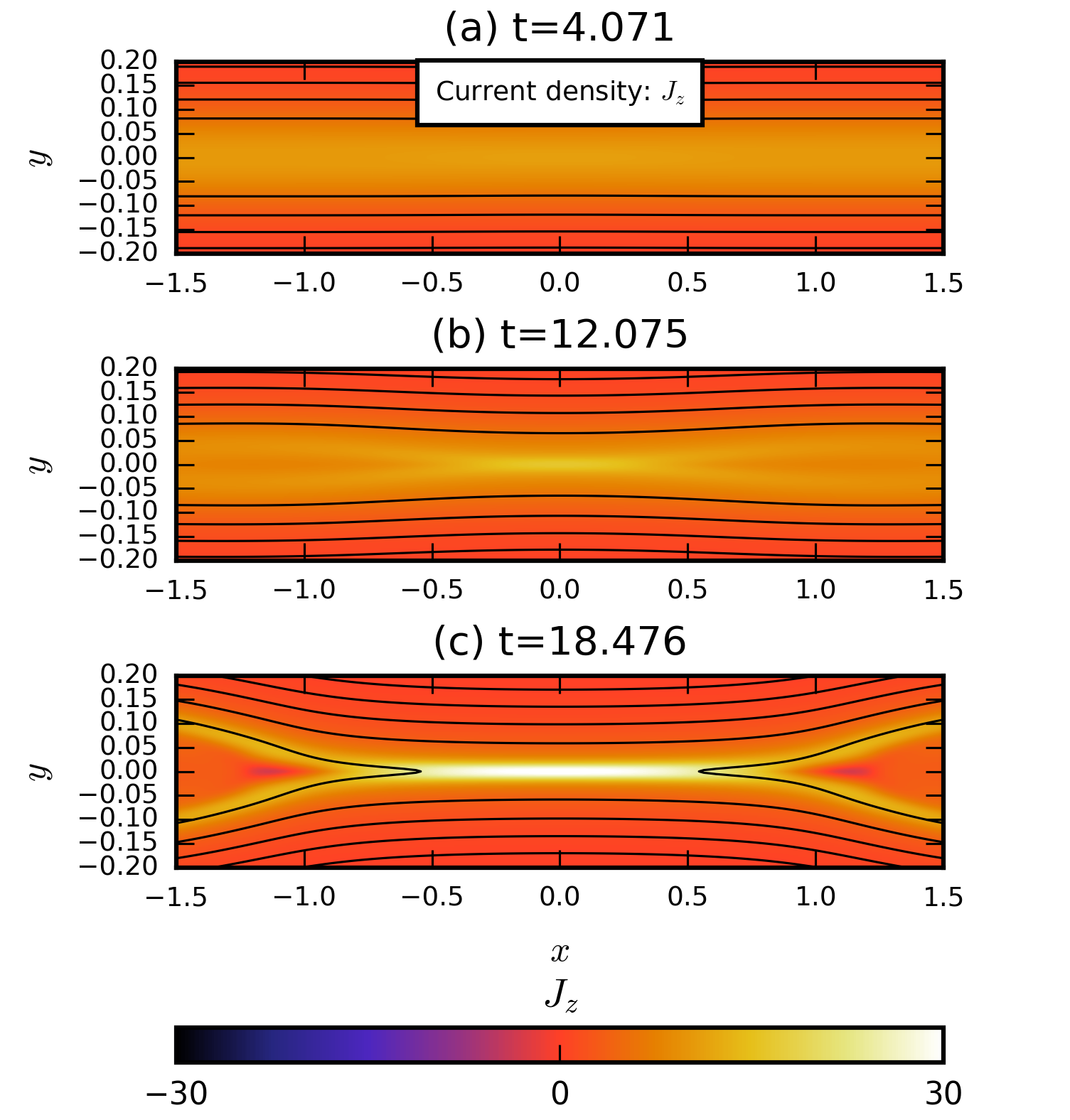}
                           \includegraphics[width=0.33\textwidth, clip=]{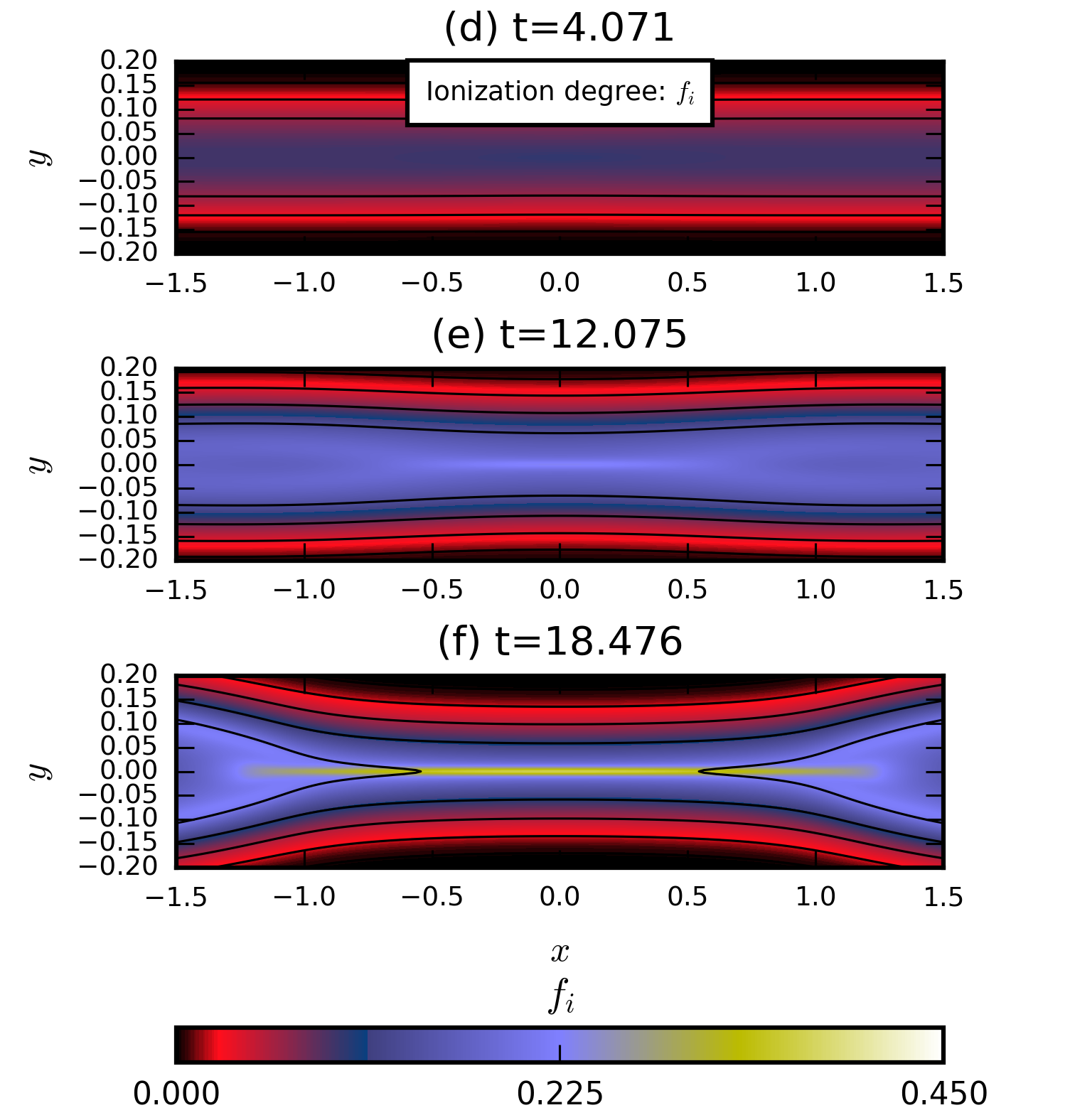}
                           \includegraphics[width=0.33\textwidth, clip=]{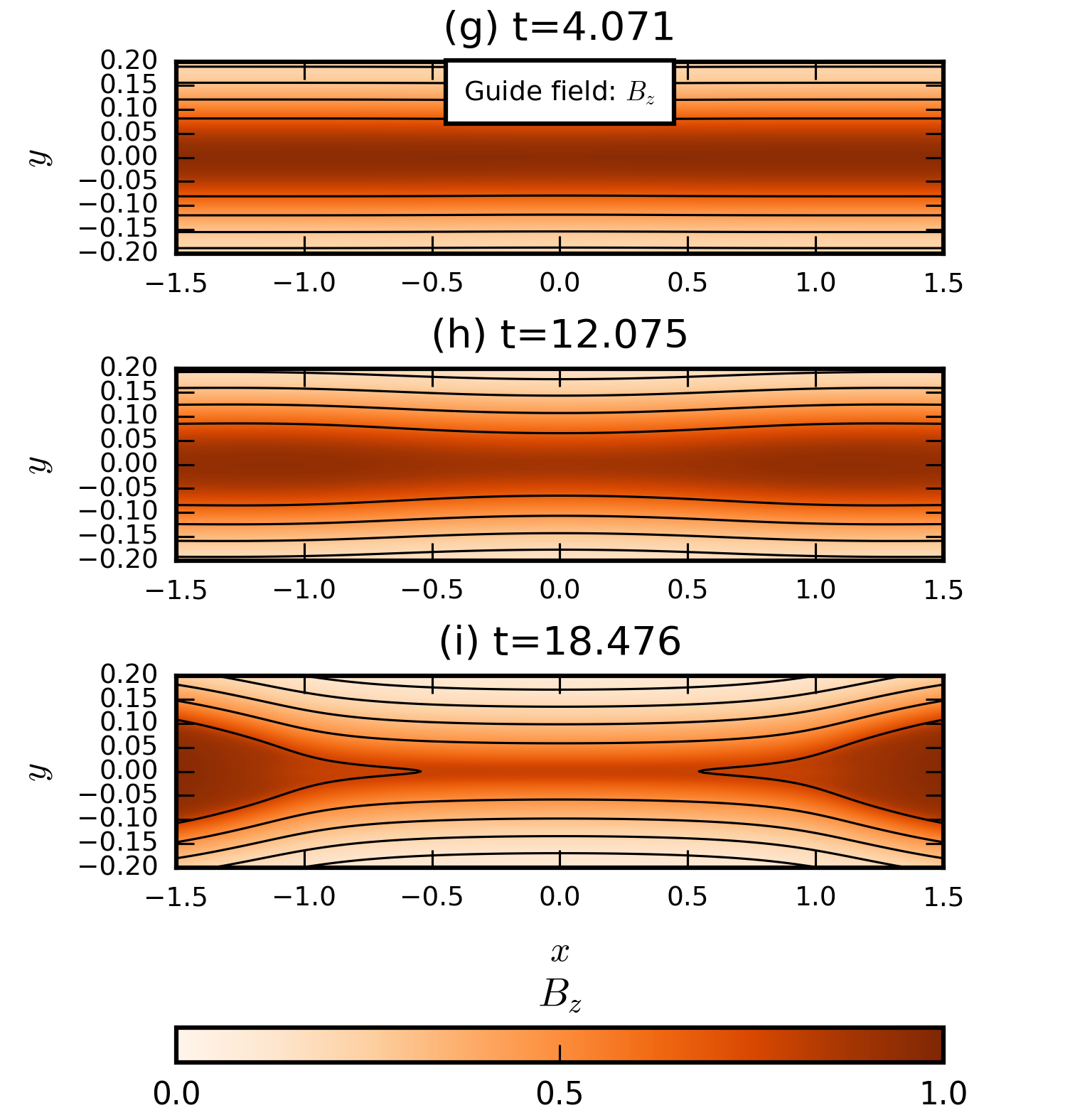} }
      \caption{Panels (a), (b) and (c) are for the current density $J_z$ in the full domain at $t=4.071$, $t=12.075$ and $t=18.476$ in Case~B; those in (d), (e) and (f) for the ionization degree $f_i$; and (g), (h) and (i) display the guide field $B_z$ .}
    \label{fig.8}
\end{figure} 
  
\begin{figure}
     \centerline{\includegraphics[width=0.33\textwidth, clip=]{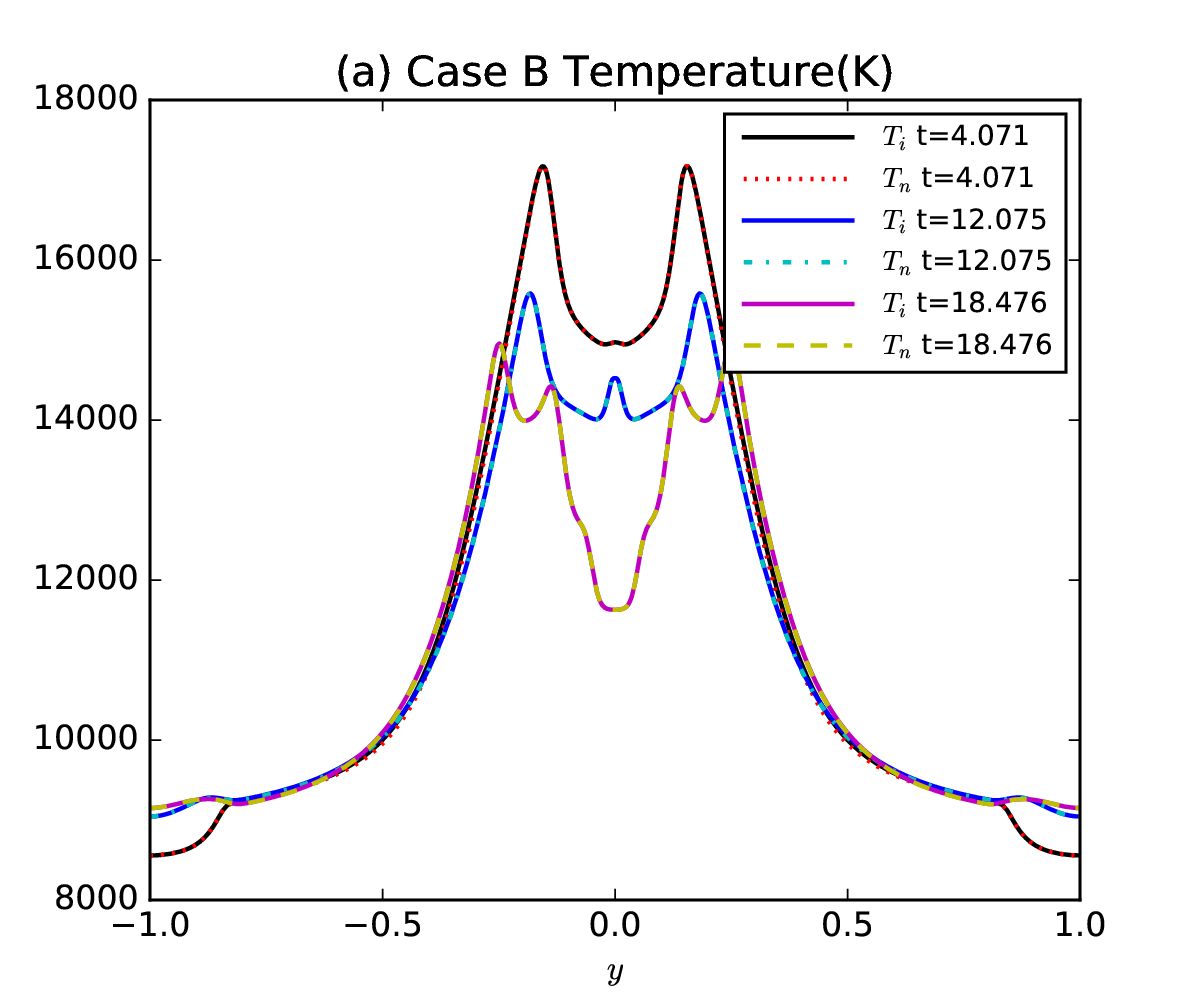}
                           \includegraphics[width=0.33\textwidth, clip=]{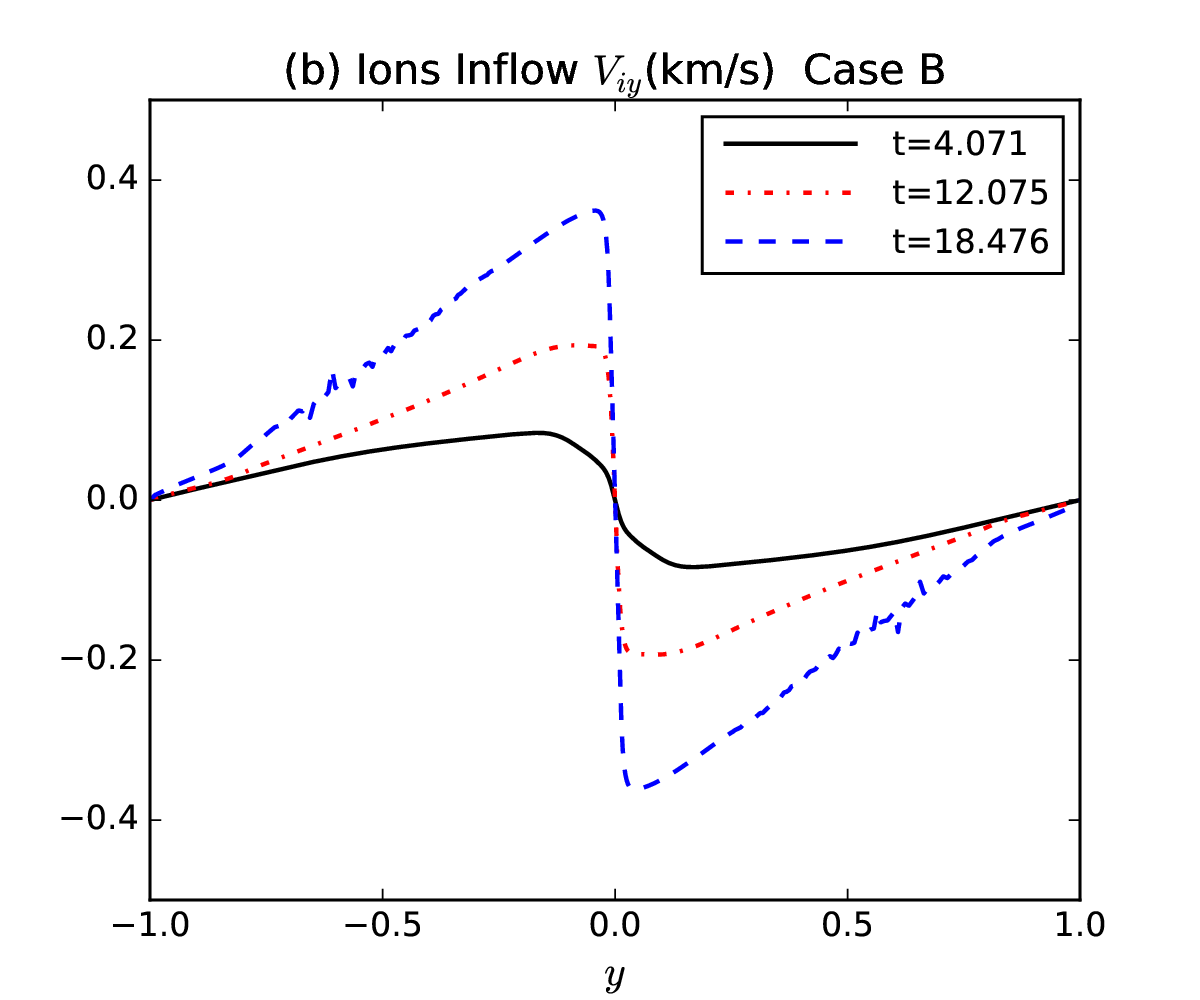}
                          \includegraphics[width=0.33\textwidth, clip=]{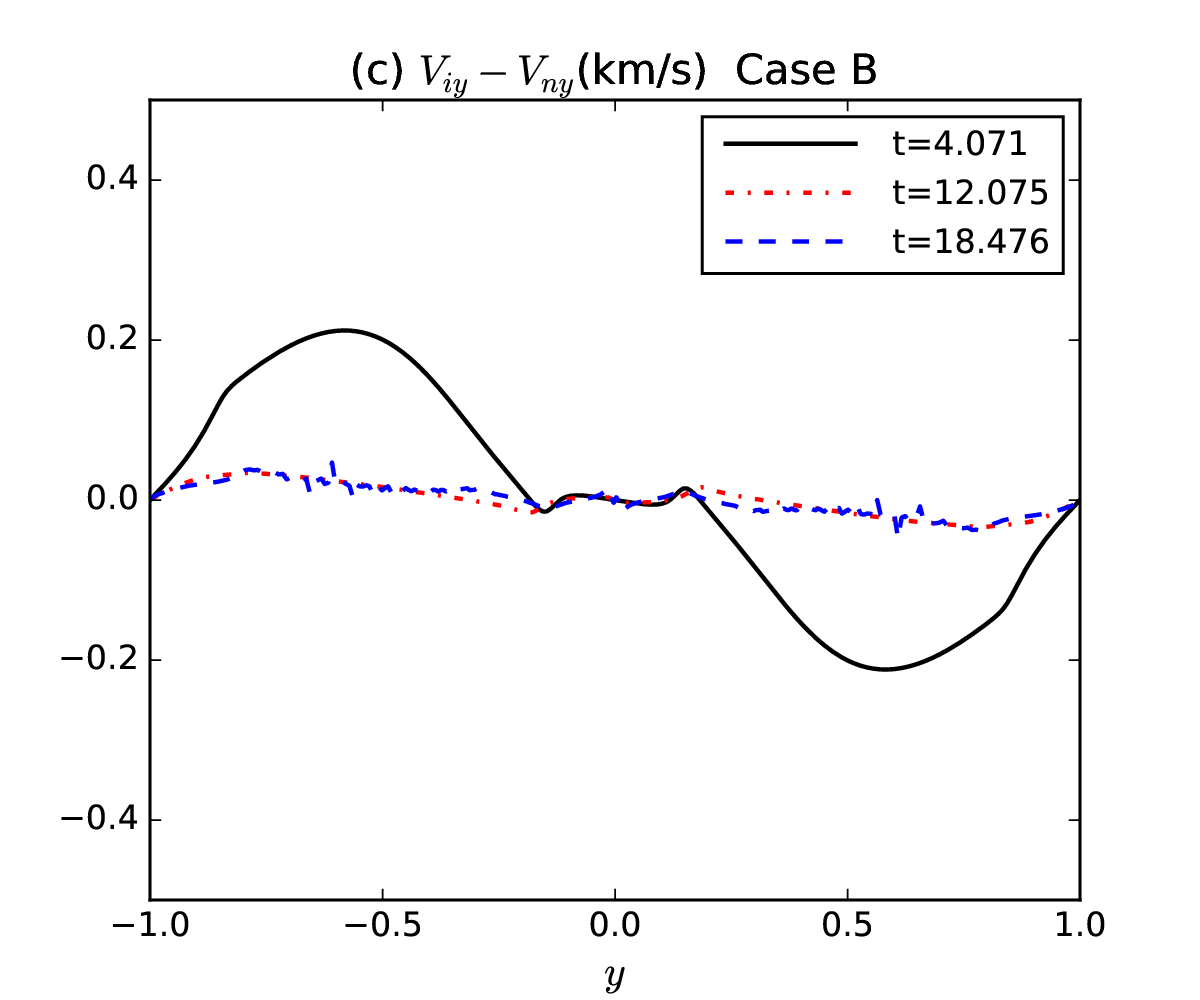} }
      \caption{(a) Distributions of the ion temperature $T_i$ and neutral temperature $T_n$ in Kelvin at $x=0$ along the $y$ direction at $t=4.071$, $t=12.075$ and $t=18.476$ in Case~B; (b) the ion inflow $V_{iy}$ in km/s at $x=0$ along the $y$ direction at $t=4.071$, $t=12.075$ and $t=18.476$ in Case~B; and (c) the difference in speed between the ion and the neutral inflows $V_{iy}-V_{ny}$ in km/s at $x=0$ along the $y$ direction at $t=4.071$, $t=12.075$ and $t=18.476$ in Case~B.}
     \label{fig.9}
 \end{figure}

---------------------------

\end{document}